\documentclass[11pt]{article}
\usepackage{graphicx}
\usepackage{epstopdf}
\usepackage{amsmath}
\usepackage{amssymb}
\usepackage{amsfonts}
\usepackage{amsthm}
\usepackage[usenames]{color}
\usepackage{array}
\usepackage[
      colorlinks=true,
      linkcolor=blue,
      urlcolor=blue,
      filecolor=blue,
      citecolor=red,
      pdfstartview=FitV,
      pdftitle={},
      pdfauthor={},
      pdfsubject={},
      pdfkeywords={},
      pdfpagemode=None,
      bookmarksopen=true
]{hyperref}

\usepackage{epsfig}

\usepackage{hyperref}
\def\fft#1#2{{#1 \over #2}}

\def\half{\frac{1}{2}}

\def\beq{\begin{eqnarray}}
\def\eeq{\end{eqnarray}}

\def\l{\lambda}
\def\La{\Lambda}

\def\O{\Omega}

\def\th{\theta}

\def\be{\begin{equation}}
\def\ee{\end{equation}}
\def\bea{\begin{eqnarray}}
\def\eea{\end{eqnarray}}

\def\be{\begin{equation}}
\def\ee{\end{equation}}
\def\bea{\begin{eqnarray}}
\def\eea{\end{eqnarray}}

\newcommand{\rom}[1]{\mathrm{#1}}

\def\cG{\mathcal{G}}

\def\cI{\mathcal{I}}

\def\cR{\mathcal{R}}

\def\nn{\nonumber}

\numberwithin{equation}{section}

\begin{document}

\begin{centering}

\textbf{\LARGE{Conserved charges in asymptotically \\   \vspace{0.2cm} de Sitter spacetimes}}

 \vspace{0.8cm}
 P B Aneesh, Sk Jahanur Hoque, 
Amitabh Virmani

  \vspace{0.5cm}

 \vspace{0.5cm}

\begin{minipage}{.9\textwidth}\small  \begin{center}
Chennai Mathematical Institute, H1 SIPCOT IT Park, \\ Kelambakkam, Tamil Nadu, India 603103\\
  \vspace{0.5cm}
{\tt  aneeshpb, skjhoque, avirmani@cmi.ac.in}
\\ $ \, $ \\
\end{center}
\end{minipage}

\end{centering}

\begin{abstract}
We present a covariant phase space construction of hamiltonian generators of asymptotic symmetries with  ``Dirichlet'' boundary conditions in de Sitter spacetime, extending a previous study of J\"ager. We show that the de Sitter charges so defined are identical to those of Ashtekar, Bonga, and Kesavan (ABK). 
We then present a comparison of ABK charges with other notions of de Sitter charges. We compare ABK charges with counterterm charges, showing that they differ only by a constant offset, which is determined in terms of the  boundary metric alone. We also compare ABK charges with charges defined by Kelly and Marolf at spatial infinity of de sitter spacetime. When the formalisms can be compared, we show that the two definitions agree. Finally, we express Kerr-de Sitter metrics in four and five dimensions in an appropriate Fefferman-Graham form.
\end{abstract}

\newpage

\tableofcontents

\section{Introduction and summary of results}
The success of the inflationary paradigm in the early universe cosmology and the discovery of the accelerated expansion of the universe in the current epoch provide ample motivation to understand  de Sitter and asymptotically de Sitter spacetimes. In contrast, from a theoretical point of view, de Sitter spacetime poses numerous challenges. The existence of de Sitter solutions in string theory and finiteness of the entropy of de Sitter horizon are some of the most debated topics in modern theoretical physics. Even classical issues in gravitational physics, such as the notion of gravitational waves in de Sitter, phase space of asymptotically de Sitter spacetimes, and appropriate notions of conserved charges in asymptotically de Sitter spacetime, are less well explored  than for asymptotically flat or asymptotically anti-de Sitter settings.

Our work explores notions of asymptotically de Sitter spacetimes and associated  conserved charges. Over the last two decades there have been many discussions of asymptotically de Sitter spacetimes and de Sitter charges, see, e.g.,~\cite{Abbott:1981ff, Strominger:2001pn, Balasubramanian:2001nb,  Kastor:2002fu, Jaeger, Anninos:2010zf, Kelly:2012zc, ABKI, Poole:2018koa} and reviews
~\cite{Anninos:2012qw, Ashtekar:2017dlf, Szabados:2018erf} for further references.
A well studied notion of asymptotically de Sitter spacetimes in non-linear general relativity is ``Dirichlet'' or ``reflective'' boundary conditions at future infinity $\cal{I}^+$.

At the outset, let us point out that Dirichlet boundary conditions are not fully satisfactory for various reasons. Firstly,  these boundary conditions rule out fluxes of de Sitter charges across $\cal{I}^+$. Thus, with these boundary conditions, gravitational waves do not carry away de Sitter charges, say, mass and momentum, across $\cal{I}^+$. Secondly, by definition since the boundary metric is held fixed, symplectic structure computed on a complete Cauchy slice necessarily vanishes.\footnote{We show this in detail in section \ref{sec:charges}.} This clearly suggests that these boundary conditions are too restrictive.  Alternative boundary conditions have been proposed~\cite{Anninos:2010zf, Kelly:2012zc, Ashtekar:2017dlf}, though they remain less explored. This needs to be contrasted with the recent developments in linearised gravity in de Sitter, where now there is a good control over many calculations~\cite{ABKII, ABKIII, ABKIIIPRL,Date:2015kma,Date:2016uzr, Hoque:2018byx}.

In this work, we expand on the previous studies with Dirichlet boundary conditions in de Sitter, with the hope that it will pave a path for addressing more difficult questions. More precisely, we work with boundary conditions of Ashtekar, Bonga, and Kesavan (ABK)~\cite{ABKI}. ABK have also presented a construction of de Sitter charges via an analysis of asymptotic equations of motion in the conformal infinity framework.  From their analysis it is not clear in what sense the charges they have defined are generators of asymptotic symmetries. One aim of our work is to present a covariant phase construction of ABK charges.  Another aim is to compare ABK charges with other approaches, specifically counterterm charges~\cite{Strominger:2001pn, Balasubramanian:2001nb} and  Kelly-Marolf charges~\cite{Kelly:2012zc}.

One can give certain indirect arguments  for relating these various notions of charges~\cite{Kelly:2012zc}. However, these indirect arguments are hardly illuminating; an explicit comparison between  various approaches remain fairly cumbersome as these various approaches are based on very different techniques: ABK use conformal infinity framework whereas counterterm method uses Fefferman-Graham expansion, and Kelly-Marolf use radial expansion in ADM form near spatial infinity. A priori it is not at all obvious how different quantities appearing in the corresponding expressions of charges can be compared with each other. In this work we demystify these connections; we present a direct comparison between ABK charges and counterterm charges, and between ABK charges and  Kelly-Marolf charges.

The rest of the paper is organised as follows.  In section \ref{sec:charges}, after defining our notion of asymptotically de Sitter spacetimes,  we first provide a covariant phase space construction of conserved charges in de Sitter spacetimes using the general formalism of Wald and Zoupas~\cite{Wald:1999wa}. We show that conserved charges so defined are manifestly equivalent to  ABK 
charges. This part of our work is inspired by the corresponding AdS analysis of Hollands, Ishibashi, and Marolf~\cite{HIM}, whose technology we closely follow.  
After we worked on this idea, we came to know about the master's thesis of S.~J\"ager~\cite{Jaeger}. He also carried out covariant phase space construction similar to ours  ten years ago. Our motivation and aim was to  interpret ABK charges as hamiltonian generators of asymptotic symmetries; at the time of J\"ager's analysis, the work of ABK did not exist.  Our section \ref{asym}, section \ref{sec:WZ_ABK} and appendix \ref{App_asym} overlap with J\"ager's thesis. In section \ref{sec:counterterm} we present a direct comparison between ABK charges and counterterm charges~\cite{Strominger:2001pn, Balasubramanian:2001nb, Klemm:2001ea}. As expected from the corresponding AdS analysis~\cite{HIM}, these  two approaches are not equivalent. However, as is well known in the AdS context, the difference is only a constant offset, expressible in terms of non-dynamical boundary data. Analogous results are obtained for de Sitter spacetimes. As a result of a detailed computation we also show that the trace of the counterterm stress tensor precisely matches with the trace-anomaly computed in~\cite{Nojiri:2001mf}.

In section \ref{sec:MK_ABK} we  compare  ABK charges to Kelly-Marolf charges. The Kelly-Marolf definition is conceptually very different from ABK definition. We very briefly review salient features of Kelly-Marolf construction, and argue that at least for a class of spacetimes  Kelly-Marolf  boundary conditions are compatible with ABK boundary conditions. For this class of spacetimes, we show that the two expressions of the charges are equivalent.  It is not clear to us for what classes of asymptotically de Sitter spacetimes Kelly-Marolf and ABK boundary conditions are compatible. We have not attempted to address this question in this work. 
  
   Finally, in section \ref{sec:examples} we express four and five dimensional Kerr-de Sitter metrics in a form that manifests the fact that they belong to our phase space.   A number of appendices complete the technical aspects of our analysis.  For the Riemann tensor (both bulk and boundary) our conventions are same as Wald's textbook~\cite{Wald:1984rg}, $[\nabla_a, \nabla_b] k_c = R_{abc}{}^{d} k_d,$ $R_{ac} = R_{abc}{}^{b}$ and $R = R_{ab} g^{ab}$.

\section{de Sitter charges at scri}
\label{sec:charges}
In  section \ref{sec:asym_dS} we start with our definition of asymptotically  de Sitter spacetimes. In section \ref{asym} we present a summary of the analysis of appendix \ref{App_asym}   of the asymptotic equations of motion near ${\cal I}^+$. In section \ref{sec:WZ_ABK} after a  brief review of the general formalism of Wald and Zoupas~\cite{Wald:1999wa}, we present a construction of conserved charges with our notion of asymptotically de Sitter spacetimes.  Comparison to counterterm charges is presented in section \ref{sec:counterterm}.

\subsection{Asymptotically  de Sitter spacetimes}
\label{sec:asym_dS}

Following~\cite{ABKI} we define  (future)  asymptotically de Sitter spacetimes as follows. A spacetime $\{ {\cal M},  g_{ab} \}$ satisfying Einstein equations with positive cosmological constant $\Lambda >0$,
\be
R_{ab} - \frac{1}{2} R g_{ab} + \Lambda g_{ab} = 0,
\ee
is asymptotically de Sitter if there exist a manifold $\widetilde{\cal{M}}$ with boundary ${\cal I}^+$  with metric $\tilde g_{ab}$ and a diffeomorphism from the interior of $\widetilde{\cal{M}}$ to  $\cal M$ such that:
\begin{enumerate}
\item There is a smooth function $\Omega$ on $\widetilde{\cal{M}}$ 
with the properties: $(i)$ $\Omega = 0 $ on ${\cal I}^+$, $(ii)$ $\nabla_a \Omega$ is nowhere vanishing on ${\cal I}^+$, $(iii)$ the \emph{unphysical} metric $\tilde g_{ab}$ is related to the physical metric via
 $\tilde g_{ab} = \Omega^2 g_{ab}$.
\item The induced metric on ${\cal I}^+$ is locally isometric to the round metric on $(d-1)$ unit sphere $\mathbb{S}^{d-1}$. 
\end{enumerate}

The boundary ${\cal I}^+$ is space-like.  For globally asymptotically de Sitter spacetimes the boundary ${\cal I}^+$ has the topology of $\mathbb{S}^{d-1}$ sphere. For  cosmological applications we work with asymptotically de Sitter spacetimes in the Poincar\'e patch, where the boundary ${\cal I}^+$ has the topology of  $\mathbb{R}^{d-1} \simeq \mathbb{S}^{d-1}/\{p\} $. For discussions pertaining to black holes, we take boundary ${\cal I}^+$ to have the topology of   $ \mathbb{R}~\times~\mathbb{S}^{d-2} \simeq \mathbb{S}^{d-1}/\{p_1, p_2\}$.

The above definition only fixes the conformal factor $\Omega$ at ${\cal I}^+$. It is possible to choose the conformal factor in such a way that the unphysical metric in a neighbourhood of ${\cal I}^+$ takes the form~\cite{HIM} , 
\be
\tilde g_{ab} = - \tilde \nabla_a \Omega \tilde \nabla_b \Omega + \tilde h_{ab}(\Omega). \label{unphys_GN}
\ee
For the most part we will work with  form \eqref{unphys_GN} of the metric. 

To familiarise the reader with our notation, let us write pure de Sitter  metric in a form we would like to work with.
 de Sitter metric in a familiar set of global coordinates takes the form,
\be
ds^2 = - d\tau^2 + ( \cosh \tau)^2  d\sigma^2_{d-1},
\ee
where $d\sigma^2_{d-1}$ is the round metric on $(d-1)$ unit sphere. 
For simplicity  we have set the de Sitter length $\ell = 1$; de Sitter length is defined via $\ell =\sqrt{\frac{(d-1)(d-2)}{2\Lambda}}$. 
 This metric is in Gaussian normal form. We want the unphysical metric to be in Gaussian normal form~\eqref{unphys_GN}, so we define, 
\be
\tau = - \ln \left(\frac{1}{2}\Omega\right).
\ee
For the physical metric we get, 
\be
ds^2 =  \frac{1}{\Omega^2}\left[ - d\Omega^2 + \left(1 + \frac{1}{2}\Omega^2 + \frac{1}{16}\Omega^4 \right)  d\sigma^2_{d-1} \right].
\ee
and for the unphysical metric we get,
\be
\widetilde{(ds)}^2  = \Omega^{2} ds^2  =- d\Omega^2 + \left(1 + \frac{1}{2}\Omega^2 + \frac{1}{16}\Omega^4 \right)  d\sigma^2_{d-1}. 
\ee
The unphysical metric is of the form~\eqref{unphys_GN} with,
\be
\tilde h_{ab} (\Omega) = \left(1 + \frac{1}{2}\Omega^2 + \frac{1}{16}\Omega^4 \right) (\tilde h_{ab})_0,
\ee
where $(\tilde h_{ab})_0$ is the round metric on the unit $(d-1)$-sphere.

ABK also analysed asymptotic symmetries for the above definition of asymptotically de Sitter spacetimes. They have shown that the asymptotic symmetries are just the conformal isometries of the boundary metric. For $\mathbb{S}^{d-1}$ topology, this group is simply the de Sitter group $\mathrm{SO}(1,d)$. Translational isometries of de Sitter space are represented as conformal Killing vectors of the sphere and rotational isometries of de Sitter are represented as  Killing vectors of the sphere. For  $\mathbb{R}^{d-1}$ and $\mathbb{R}\times\mathbb{S}^{d-2}$  boundary topologies, ABK present a detailed discussion to which we refer the reader.

\subsection{Asymptotic expansion}
\label{asym}
A detailed  analysis of Einstein equations is required near ${\cal I}^+$ to further characterise asymptotically de Sitter metrics.  Such an analysis is presented in appendix \ref{App_asym} and was first carried out in~\cite{Jaeger}. The salient features are as follows.  The ADM decomposition of Einstein equations for the unphysical metric with respect to $\O = \mbox{const}$ slices in gauge \eqref{unphys_GN} gives two constraint equations,
\bea
 \tilde{\cR} + \tilde{K}^2 - \tilde{K}_{ab} \tilde{K}^{ab} &=& 2 (d-2) \O^{-1} \tilde{K} \\
\tilde{D}_c \tilde{K}^c_{\ a} - \tilde{D}_a \tilde{K} &=& 0,
\eea
and two evolution equations, 
\bea 
\pounds_n \tilde{h}_{ab} &=& -2\tilde{K}_{ab}, \\
\pounds_n \tilde{K}_a^{\ b} &= & \tilde{\cR}_a^{\ b} + \tilde{K} \tilde{K}_a^{\ b} - \O^{-1} \tilde{K} \tilde{h}_a^{\ b} - (d-2) \O^{-1} \tilde{K}_a^{\ b} \label{evo_K_MT},
\eea 
where $\tilde \cR_{ab}$ and $\tilde \cR$ respectively  denote the Ricci tensor and  Ricci scalar of  metric $\tilde h_{ab}(\Omega)$, $\tilde K_{ab}$ is the unphysical extrinsic curvature of $\O = \mbox{const}$ surfaces. In our conventions\footnote{This convention appears to be standard in de Sitter literature, which differs from Wald's textbook~\cite{Wald} convention by an overall minus sign.},
\be
\tilde K_{ab} = - \tilde \nabla_a \tilde n_b 
\ee
where
\be
\tilde n_b = \tilde \nabla_b \Omega, 
\ee
is the unit normalised future directed normal to $\O = \mbox{const}$ surface with respect to the unphysical metric. $\tilde{D}_a$ is the metric compatible derivative with respect to $\tilde h_{ab}(\Omega)$ and $\tilde \nabla_a$ is the metric compatible derivative with respect  to the unphysical metric $\tilde g_{ab}$.
 It is convenient to write equation \eqref{evo_K_MT}  in terms of its trace and trace-free parts separately.  Defining, 
\be
\tilde{p}_a{}^{b} = \tilde{K}_a{}^{b} - \frac{\tilde h_a{}^b}{(d-1)} \tilde{K},
\ee
and using \be \pounds_n \equiv \frac{d}{d\O}, \ee we have
\bea
&& \frac{d}{d \O} \ \tilde{p}_a^{\ b} ~=~ - \left[ \tilde{\cR}_a^{\ b} - \frac{\tilde h_a{}^b}{(d-1)} \tilde{\cR} \right] - \tilde{K} \tilde{p}_a{}^{b} + (d-2) \O^{-1} \tilde{p}_a{}^{b} , \label{evo1_MT}\\
&& \frac{d}{d \O} \ \tilde{K} ~=~ - \tilde{R} - \tilde{K}^2 + (2d-3)\O^{-1} \tilde{K}, \label{evo2_MT}\\
&& \frac{d}{d \O} \ \tilde{h}_{ab} ~=~  2 \tilde{h}_{bc}\tilde{K}^c_{\ a} \label{evo3_MT}.
\eea

Asymptotic expansion of these equations is obtained upon substituting a Taylor expansion for the metric in powers of $\O$ ,
\be
 \tilde{h}_{ab} = \sum_{j=0}^{\infty} (\tilde{h}_{ab})_j \O^j. \label{h_Taylor}
 \ee
This expansion is accompanied with similar expansions for other quantities,
\begin{align}
&\tilde{p}_a^{\ b} = \sum_{j=0}^{\infty} (\tilde{p}_a^{\ b})_j \O^j, & 
&\tilde{K} = \sum_{j=0}^{\infty} (\tilde{K})_j \O^j, & \label{K_Taylor} \\
&\tilde{R}_{ab} = \sum_{j=0}^{\infty} (\tilde{R}_{ab})_j \O^j, &
&\tilde{R} = \sum_{j=0}^{\infty} (\tilde{R})_j \O^j. & \label{R_Taylor}
\end{align}
Inserting these expansions in equations \eqref{evo1_MT}-\eqref{evo3_MT} we get
\be
(-d+2+j) ( \tilde p_{a}{}^b)_j = - \left[ (\tilde{\cal R}_a{}^{b} )_{j-1} - \frac{\tilde h_a{}^{b}}{d-1}(\tilde {\cal  R} )_{j-1} \right]  - \sum_{m=0}^{j-1} (\tilde K)_m  ( \tilde p_{a}{}^b)_{j-1-m},
\label{FG1}
\ee
\be
(-2d + 3 + j) (\tilde K)_j =  - (\tilde{ \cal  R} )_{j-1} - \sum_{m=0}^{j-1} (\tilde K)_m  ( \tilde K)_{j-1-m},
\label{FG2}
\ee
\be
j (\tilde h_{ab})_j = 2 \sum_{m=0}^{j-1} \left[ ( \tilde h_{bc})_m (\tilde p_{a}{}^c)_{j-1-m}  + \frac{1}{d-1} (\tilde h_{ab})_m (\tilde K)_{j-1-m}\right].
\label{FG3}
\ee

These equations are an appropriate Fefferman-Graham expansion of Einstein equations for the unphysical metric. Given $( \tilde p_{a}{}^b)_0$, $(\tilde K)_0$ and $( \tilde h_{ab})_0$ these equations uniquely determine $( \tilde K_{ab})_j$ and 
$( \tilde h_{ab})_l$ for $j < (d-2)$ and $l < (d-1)$. At $j = (d-2)$ coefficient of $ (\tilde p_{a}{}^b)_{d-2}$ on the left hand side in equation \eqref{FG1} becomes zero, so $ (\tilde p_{a}{}^b)_{d-2}$ cannot be determined using these equations. Since $ (\tilde p_{a}{}^b)_{d-2}$ feeds into the right hand side of equation \eqref{FG3}, $(\tilde h_{ab})_{d-1}$ cannot be determined.  The initial data for these equations, namely, $( \tilde p_{a}{}^b)_0$, $(\tilde K)_0$ and $( \tilde h_{ab})_0$, are to be found in our definition of asymptotically de Sitter spacetimes. Multiplying equation \eqref{evo_K_MT} with $\O$ and evaluating it at $\O = 0$ it follows that $( \tilde p_{a}{}^b)_0=0$, $(\tilde K)_0=0$. In our definition of asymptotically de Sitter spacetimes, $( \tilde h_{ab})_0$ is taken to be locally isometric to the round metric on $\mathbb{S}^{d-1}$. 

The recursions \eqref{FG1}-\eqref{FG3} can continue to higher orders provided  $(\tilde p_{a}{}^b)_{d-2}$ is known. Therefore, this tensor contains all the information about the spacetime that is not contained in the definition of asymptotically de Sitter spacetimes.  As detailed in appendix \ref{App_asym}, it turns out that $(\tilde p_{a}{}^b)_{d-2}$ and hence $(\tilde h_{ab})_{d-1}$ are directly related to the electric part of the Weyl tensor. We define the unphysical electric part of the Weyl tensor as,
\be
\tilde{E}_{ac} = \frac{1}{d-3} \O^{3-d} \left( \tilde{C}_{abcd} \tilde{n}^b \tilde{n}^d \right). \label{elec_weyl}
\ee
Using ADM decomposition it follows that,
\be
\tilde{C}_{abcd} \tilde{n}^b \tilde{n}^d = \pounds_n \tilde{K}_{ac} + \tilde{K}_a^{\ b}\tilde{K}_{bc} + \O^{-1} \tilde{K}_{ac}.
\ee
A similar Taylor expansion as above gives,
\be
(\tilde{E}_{ac})_0 = \frac{1}{d-3}  \left( \tilde{C}_{abcd} \tilde{n}^b \tilde{n}^d \right)_{d-3} = - \left( \tilde{K}_{ac} \right)_{d-2} + \sum_{m=0}^{d-3} \frac{1}{d-3} \left(\tilde{K}_a^{\ b} \right)_m \left(\tilde{K}_{bc} \right)_{d-3-m}.
\ee

From this equation it follows that (for details see appendix \ref{App_asym}) in four dimensions, asymptotically de Sitter metrics take the form,

\begin{align}
\tilde g_{ab} = \begin{cases}
- \tilde \nabla_a \Omega \tilde \nabla_b \Omega + \left( 1+ \frac{1}{2} \Omega^2 \right)  (\tilde h_{ab})_0  - \frac{2}{3} \Omega^3 \tilde E_{ab} + \mathcal{O}(\Omega^{4}) & d=4,  \\
- \tilde \nabla_a \Omega \tilde \nabla_b \Omega + \left( 1+ \frac{1}{2} \Omega^2 + \frac{1}{16}\Omega^4 \right)  (\tilde h_{ab})_0  - \frac{1}{2} \Omega^4 \tilde E_{ab} + \mathcal{O}(\Omega^{5}) & d = 5 , \\
 - \tilde \nabla_a \Omega \tilde \nabla_b \Omega + \left( 1+ \frac{1}{2} \Omega^2 + \frac{1}{16}\Omega^4 \right)  (\tilde h_{ab})_0  - \frac{2}{d-1} \Omega^{d-1} \tilde E_{ab} + \mathcal{O}(\Omega^{d}) & d \ge 6 ,
 \end{cases} \label{FG_4d_MT}
\end{align}
where  $(\tilde h_{ab})_0$ is the round metric on the unit $(d-1)$-sphere.
From these expressions it is clear that in all dimensions, a class of variations that preserve our notion of  asymptotically de Sitter  spacetimes take the form
\be
\delta \tilde g_{ab} = - \frac{2}{d-1}\Omega^{d-1} (\delta \tilde E_{ab}) + \mathcal{O}(\Omega^{d}). \label{gamma_unp}
\ee
We can go from $\delta \tilde g_{ab}$ to $\delta  g_{ab}$ by simply multiplying with $\Omega^{-2}$~\cite{HIM}\footnote{This is because in the gauge fixed form \eqref{unphys_GN} of  the unphysical metrics,  $\O$ is to be thought of as a fixed function on $\widetilde {\cal M}$. It is  part of the background structure used in specifying the asymptotic conditions.}:
\be
\delta  g_{ab} = - \frac{2}{d-1}\Omega^{d-3} (\delta \tilde E_{ab}) + \mathcal{O}(\Omega^{d-2}). \label{gamma}
\ee

\subsection{Noether charges and  ABK charges}
\label{sec:WZ_ABK}
Wald and Zoupas~\cite{Wald:1999wa} have given a general formalism to 
construct conserved quantities within the covariant phase space framework~\cite{Ashtekar:1990gc, Lee:1990nz, Iyer:1994ys}, for a recent review and further references see~\cite{Compere:2018aar}. In this section we apply these ideas to asymptotically de Sitter  spacetimes characterised by the asymptotic expansion of the previous subsection.  Let 
$L$ be a diffeomorphism invariant $d$-form Lagrangian density. The variation 
of $L$ can be written as,
\be
\delta L = E(g)_{ab} \delta g^{ab} + d \theta (g,\delta g),
\ee
where $\theta$ is the presymplectic potential $(d-1)$-form. The 
Euler-Lagrange equations of motion of the theory are given by $E(g)_{ab}=0$. Now, 
consider a  two parameter family $g(\lambda_1, \lambda_2)$ of field 
configurations and  let 
\be
\delta_1 g:=\frac{\partial g}{\partial \lambda_1} \Bigg{|}_{\l_1=0,~\l_2 = 0}, \qquad \qquad 
\delta_2 g:=\frac{\partial g}{\partial \lambda_2} \Bigg{|}_{\l_1=0,~\l_2 = 0}
\ee 
be variations of $g$. The variations $\delta_1 g $ and $
\delta_2 g$ are to be thought of as tangent vectors to field configuration 
space ${\cal F}$ along the flow generated by parameters $\lambda_1$ and $
\lambda_2$ respectively.
From the presymplectic potential $\theta$, we can obtain the presymplectic current $
(d-1)$-form $\omega$ as,
\be
\omega(g, \delta_1g, \delta_2g )= \delta_1 \theta (g, \delta_2 g) - \delta_2 \theta(g, \delta_1g).
\ee
Integrating presymplectic current over a Cauchy
hypersurface\footnote{The integrals are defined with appropriate boundary conditions for fields to ensure that the integral is finite.} we obtain a presymplectic form,
\be
\Omega(g, \delta_1g , \delta_2 g) = \int_\Sigma \omega(g, \delta_1g, 
\delta_2g ).
\ee
In general, the presymplectic form $\Omega(g, \delta_1g , \delta_2 g) $ is degenerate. Details on the construction of a non-degenerate symplectic structure on the phase space can be found in~\cite{Lee:1990nz}. For our purposes,  the presymplectic form is sufficient.

We denote by $\bar{{\cal F}}$ the subspace of ${\cal F}$ whose elements are solutions to equations of motion $E(g)_{ab} = 0$.  Now, a 
vector field $\xi$ on space-time manifold $M$ with metric $g$ (a point on $\bar{{\cal F}}$) naturally induces the field variation $\delta_\xi g = \pounds_{\xi} g 
$ on $\bar{{\cal F}}$. The Hamiltonian function $H_{\xi}$   conjugate to $\xi$ is defined to be
\be \label{Var_H}
\delta H_\xi = \Omega(g, \delta g , \pounds_\xi g)=\int_\Sigma \omega(g, \delta g,  \pounds_\xi g ).
\ee
 As emphasized in~\cite{Wald:1999wa}, equation \eqref{Var_H} does 
not ensure the existence of a Hamiltonian function $H_\xi$ conjugate to $\xi^a$. To analyse this, let us define 
the Noether current $(d-1)$-form $J_{\xi}$ associated with $\xi^a$, 
\be \label{NoetherCurrent}
J_\xi = \theta(g, \pounds_\xi g) - \xi \cdot L,
\ee
where $\xi \cdot X$ denotes contraction $\xi^a$ into the first index of the form $X$. A simple calculation shows $dJ_{\xi}=-E\pounds_{\xi} g$, i.e., $J_{\xi}$ is closed 
on-shell. It can also be shown that $J_{\xi}$ is not only 
closed  but also exact~\cite{Wald:1990}. Hence we define Noether charge $(d-2)$-form $Q_{\xi}$, such that
\be \label{NoetherCharge}
J_\xi = d Q_\xi.
\ee
Taking an on-shell  variation of the Noether current $J_{\xi}$ and using equations of motion it can be shown that,
\be
\delta H_\xi =\int_{\partial \Sigma} \left[ \delta Q_\xi - \xi \cdot \theta \right]. \label{integralH}
\ee
This equation  gives rise to a necessary condition for the existence of the Hamiltonian function $H_\xi$. Considering commutator of two variations, we have
\bea
0  &=&  (\delta_1 \delta_2 - \delta_2 \delta_1) H_\xi  \\
&=& - \int_{\partial \Sigma} \xi \cdot \left[ \delta_1 \theta (g, \delta_2 g) - \delta_2 
\theta(g, \delta_1g ) \right]\\
&=& - \int_{\partial \Sigma} \xi \cdot \omega(g, \delta_1g , \delta_2 g). \label{xi_omega}
\eea
 Even though this condition seems to be only a necessary one, it turns out to be also sufficient for the existence of Hamiltonian~\cite{Wald:1999wa} with $\delta_1 g$ and $\delta_2 g $ in $\bar{{\cal F}}$. 

Now we wish to apply the above formalism to asymptotically de Sitter spacetimes. A  $d$-form Lagrangian density that yields Einstein's equation with a positive 
cosmological constant $\Lambda$ is,
\be
L_{a_1 \ldots a_d} = \frac{1}{16 \pi G} (R - 2\Lambda) \epsilon_{a_1 \ldots a_d},
\ee
where $R$ is the Ricci scalar and $\epsilon$ is the volume form associated with metric 
$g_{ab}$.  This Lagrangian gives rise to the field equations,
\be
E(g)^{ab}{}_{a_1 \ldots a_d} = \frac{1}{16 \pi G} \left(R^{ab}  - \frac{1}{2} R g^{ab} + 
\Lambda g^{ab}\right) \epsilon_{a_1 \ldots a_d},
\ee
and the presymplectic potential
\be \label{PreSymPot}
\theta_{a_1 \ldots a_{d-1}} =  \frac{1}{16 \pi G}  v^c \epsilon_{c a_1 \ldots a_{d-1}},
\ee
where
\be
v^c = g^{cd} g^{ef} ( \nabla_f \delta g_{de} - \nabla_d \delta g_{ef}).
\ee
From \eqref{PreSymPot},  the presymplectic current $(d-1)$-form can be obtained,
\be
\omega_{a_1 \ldots a_{d-1}} =  \frac{1}{16 \pi G}  \omega^c \epsilon_{c a_1 \ldots a_{d-1}} .
\ee
where 
\be
\omega^a = - P^{abcdef} (  \delta_1 g_{bc} \nabla_{d} \delta_2 g_{ef} - \delta_2 g_{bc} \nabla_{d} \delta_1 g_{ef} ), \label{omega_a}
\ee
with
\be
P^{a b c d e f} =  g^{a e}  g^{f b}  g^{c d} - \frac{1}{2}  g^{a d}  g^{b e}  g^{f c} - \frac{1}{2}  g^{a b}  g^{c d}  g^{e f} - \frac{1}{2}  g^{b c}  g^{a e}  g^{f d} + \frac{1}{2}   g^{b c}  g^{a d}  g^{e f}. \label{Pabcdef}
\ee
Noether current \eqref{NoetherCurrent}  takes the form,
\be
(J_\xi)_{a_1 \ldots a_{d-1}} = \frac{1}{8 \pi G} \nabla_c \nabla^{[c}\xi^{b]} \epsilon_{ba_1\ldots a_{d-1}},
\ee
and the corresponding Noether charge \eqref{NoetherCharge} can be taken to be,
\be
(Q_\xi)_{a_1 \ldots a_{d-2}} = -\frac{1}{16 \pi G} \nabla^{b}\xi^{c} \epsilon_{bca_1\ldots a_{d-2}}. \label{noetherQ}
\ee

The above expressions are written in terms of physical variables. In order to relate them to our definition of asymptotically de Sitter spacetimes, we need to convert the relevant expressions in terms of unphysical variables. For the Noether charge expression \eqref{noetherQ} we proceed as follows. Under conformal transformation,
\be
\nabla^{b}\xi^{c} = g^{be} \nabla_e \xi^c = \Omega^{2} \tilde g^{be} \left( \tilde \nabla_e \xi^c - \Omega^{-1} (\delta^c_e  \tilde \nabla_d \Omega + \delta^c_d \tilde \nabla_e \Omega -   \tilde g_{ed}\tilde g^{cf} \tilde \nabla_f \Omega ) \xi^d\right), 
\ee 
which gives
\be
(Q_\xi)_{a_1 \ldots a_{d-2}} = \frac{1}{8\pi G} \Omega^{1-d} \tilde \epsilon_{a_1 \ldots a_{d-2}bc} (\tilde \nabla^b \Omega)  \xi^c - \frac{1}{16\pi G} \Omega^{2-d} \tilde \epsilon_{a_1 \ldots a_{d-2}bc} \tilde g^{be} \tilde \nabla_e \xi^c.
\ee		
From this expression after a bit of calculation it follows that
\be
(\delta Q_\xi)_{a_1 \ldots a_{d-2}} = \frac{1}{8\pi G} \tilde \epsilon_{a_1 \ldots a_{d-2}bc} (\tilde \nabla^b \Omega) \delta \tilde E^c{}_d \xi^d + {\cal{O}}(\Omega),
\label{deltaQ}
\ee
for the class of variations \eqref{gamma}.

The most general variation consistent with our gauge choice and boundary condition is of the form~\cite{HIM},
 \be
\delta g_{ab} = - \frac{2}{d-1}\Omega^{d-3} \left(\delta \tilde E_{ab} + \pounds_\eta \tilde E_{ab} \right) + \mathcal{O}(\Omega^{d-2})
\ee
where $\eta$ is an arbitrary diffeomorphism with
\be
\pounds_\eta \bar g = \mathcal{O}(\Omega^{d-2}),
\ee
and where $\bar g$ is the background de Sitter metric. Since $\delta g_{ab}$ is $\mathcal{O}(\Omega^{d-3})$, we conclude that
\be
\nabla_a \delta g_{bc} = \tilde \nabla_a \delta g_{bc} + \mathcal{\mathcal{O}}(\Omega^{d-4}) \sim \mathcal{\mathcal{O}}(\Omega^{d-4}).
\ee
Therefore, from equations \eqref{omega_a} and \eqref{Pabcdef} we have
\be
\omega^a \sim  P \cdot (\delta g) \cdot \nabla \delta g \sim  \mathcal{O}( \Omega^{6} ) \cdot  \mathcal{O} (\Omega^{d-3})  \cdot \mathcal{O} (\Omega^{d-4})   = \mathcal{O} (\Omega^{2d-1}).
\ee
This implies that near $\mathcal{I}^+$,
\be
\omega_{a_1 a_2 a_{d-1}} = \epsilon_{a a_1 a_2 a_{d-1}} \omega^a \sim \mathcal{O}(\Omega^{-d})\cdot \mathcal{O} (\Omega^{2d-1}) \sim \mathcal{O} (\Omega^{d-1}),
\ee
i.e., 
\be 
\omega_{a_1 a_2 a_{d-1}} = 0 \qquad \mbox{at} \qquad  \mathcal{I}^+.
\ee
The presymplectic current vanishes at $\mathcal{I}^+$. A similar argument shows that the presymplectic potential $\theta (g, \delta g)$ also vanishes at  $\mathcal{I}^+$, see appendix \ref{App_asym}. Integrating the presymplectic current over complete  $\mathcal{I}^+$ it follows that the presymplectic two-form $\Omega(g, \delta_1 g, \delta_2 g)$  vanishes at the boundary $\mathcal{I}^+$. From conservation properties of the presymplectic current,  it follows that the \emph{presymplectic two-form $\Omega(g, \delta_1 g, \delta_2 g)$  vanishes on all complete Cauchy slices.} This clearly shows that the boundary conditions we work with are too restrictive. 

However, all is not lost. There is a still a useful (but formal) notion of the conserved quantities one can define.\footnote{Timelike directions play a preferred role in the covariant phase space discussion. However, we suspect that formally one can ``wick-rotate'' and define a non-degenerate symplectic structure in a ``radial'' direction that leads to conserved charges of the type discussed in this work.}
Instead of working with complete Cauchy slices, we restrict our analysis to spacelike hypersurfaces $\Sigma$ in the physical spacetime that extend smoothly to ${\cal I}^+$ of an unphysical spacetime such that the intersection of  $\Sigma$ and ${\cal I}^+$ is a smooth $(d-2)$ surface $C \subset {\cal I}^+ $. In the physical spacetime this is to be thought of as a limiting process, where one draws nested sequence of compact subsets of $\Sigma$ approaching $C$.  Then from the fact that the presymplectic current vanishes on  
${\cal I}^+ $, cf.~discussion around equation \eqref{xi_omega}, it follows that $H_\xi$ exists and the integral \eqref{integralH} is convergent and is independent of hypersurface approaching $C$.

\begin{figure}[t]
\begin{center}
\includegraphics[width=0.5 \textwidth]{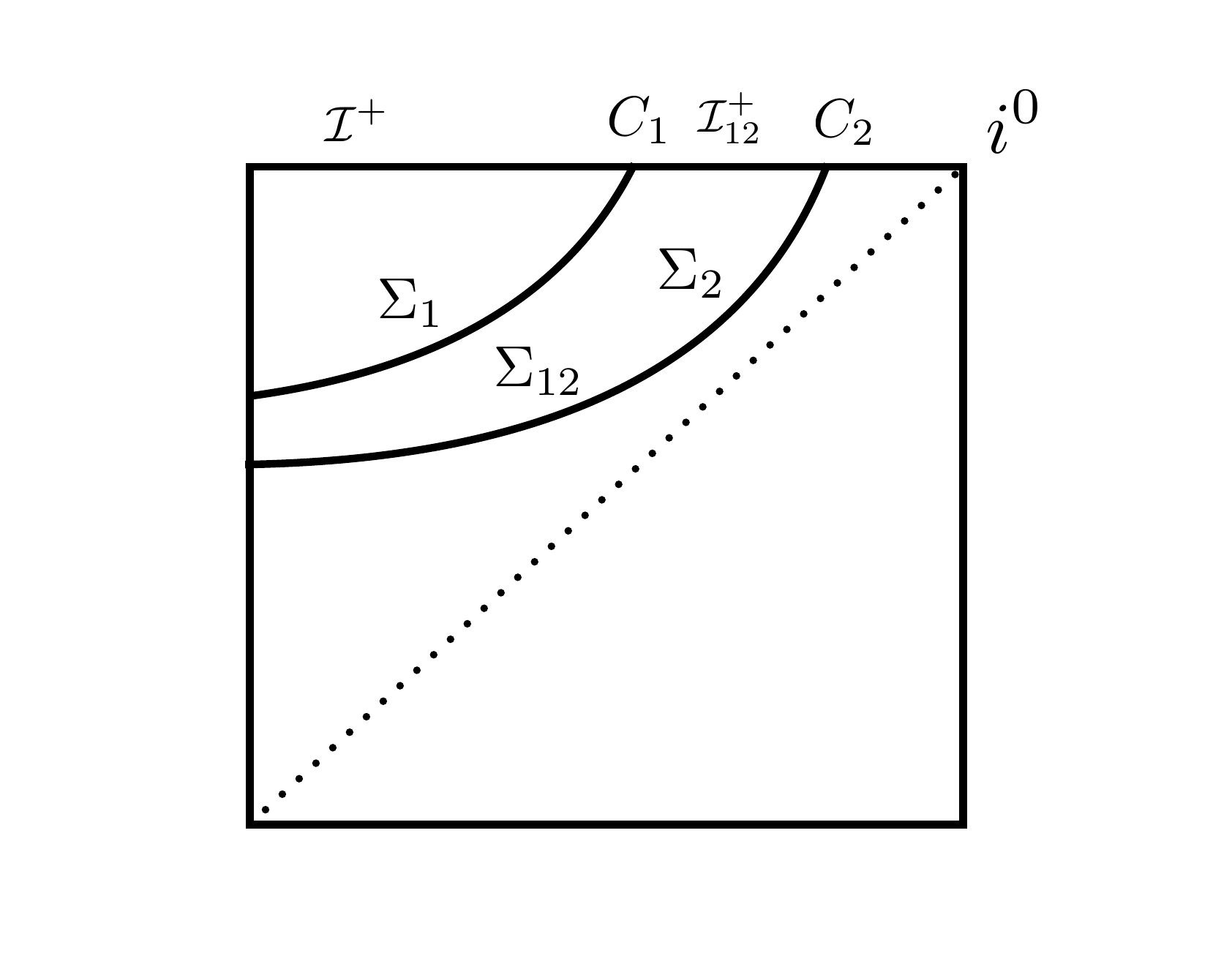}
\caption{{\sffamily
 Hypersurfaces $\Sigma_1$ and $\Sigma_2$ together with the portion ${\cal I}^+_{12}$ of ${\cal I}^+$  enclosing the spacetime volume $\Sigma_{12}$.} }
\label{figure1}
\end{center}
\end{figure}

Now that we have argued that $H_\xi$ exist, we can investigate its conservation properties. Consider two hypersurfaces $\Sigma_1$ and $\Sigma_2$ together with  a portion ${\cal I}^+_{12} $ of  ${\cal I}^+$, enclosing a spacetime volume $\Sigma_{12}$ as in figure~\ref{figure1}. The difference
\be
\delta H_\xi [\Sigma_1] - \delta H_\xi [\Sigma_2] = -\int_{{\cal I}^+_{12} } \omega(g, \delta g, \pounds_\xi g) = 0,
\ee
as $\omega(g, \delta_1 g, \delta_2 g) = 0$ on ${\cal I}^+$ . This shows that $\delta H_\xi$ in independent of the choice of the hypersurfaces as long as hypersurfaces together with a portion of ${\cal I}^+$ enclose a spacetime volume.

To further simplify and interpret expression \eqref{deltaQ}, let us recall that the volume form on the cut $C$ of the boundary ${\cal I}^+$ is related to the volume form on the unphysical spacetime as
\be
 {}^{(d)}\tilde \epsilon =\tilde u  \wedge \tilde n  \wedge
 {}^{(d-2)}\tilde \epsilon .
   \ee
Inserting this expression in  \eqref{deltaQ} and integrating over the cut $C$ we have
\be
\int_C \delta Q_\xi =  \frac{1}{8 \pi G} \ \delta \int_C \tilde E_{ab} \tilde u^b \xi^a d\tilde S.
\ee
Since the presymplectic potential vanishes at ${\cal I}^+$, we can write
\be
\delta H_\xi = \int_C \delta Q_\xi.
\ee
Taking the reference spacetime to be pure de Sitter, we have the result, 
\be \label{Charge_ABK}
H_{\xi}[C]:=\frac{1}{8 \pi G} \int _{C} \tilde E_{ab}  \tilde u^b \xi^a \widetilde{d S}.
\ee
This expression is manifestly equivalent to the corresponding ABK expression~\cite{ABKI}. As emphasised in introduction, strategy that ABK followed is very different from ours. From an analysis of asymptotic equations of motion they observed that the electric part of the Weyl tensor $\tilde E_{ab}$ is traceless and conserved at  ${\cal I}^+$. It then  follows that $\tilde E_{ab} \xi^a$ is a conserved current that allows one to define a conserved charge: $H_{\xi}[C_1] = H_{\xi}[C_2]$.

\subsection{Comparison with countertem charges for $d=5$}
\label{sec:counterterm}

There are several other definitions of charges for asymptotically de Sitter spacetime. One such approach is the AdS/CFT inspired~\cite{Balasubramanian:1999re, Emparan:1999pm, de Boer:1999xf, de Haro:2000xn, Papadimitriou:2005ii} counterterm method~\cite{Strominger:2001pn, Balasubramanian:2001nb, Klemm:2001ea}. In the counterterm method, charges associated to asymptotic symmetry $\xi^a$ are constructed from a boundary stress tensor, which is obtained by varying the effective boundary Lagrangian. The counterterm charges are defined as 
\be
Q^{\rom{ct}}_{\xi}[C] = \lim_{C_\Omega \to C} \int_{C_\Omega} \tau_{ab} \, \xi^a  \,  u^b \, dS, \label{charge_CT}
\ee
where $C_\Omega$ is a sequence of cross-sections within a partial Cauchy surface $\Sigma$ taken to  $C$ in $\cal{I}^+$ and  $u^b$ is the unit normal to $C_\Omega$ in $\Sigma$. The form of the boundary stress tensor $\tau_{ab}$ depends on the number of dimensions. Following~\cite{AshtekarDas, HIM} we restrict our attention to five-dimensions, though some expressions we write with explicit $d$,
\be
{\tau}_{ab} 
=\frac{1}{8\pi G}\bigg[({K}{h}_{ab}-{K}_{ab})+(d-2){h}_{ab}+\frac{1}
{d-3}\left(\cR_{ab}-\frac{1}{2} \cR {h}_{ab}\right)\bigg].
\ee
To avoid any potential confusion, as different papers use different conventions, we recall that our conventions are 
\be
K_{ab} = - h_a{}^{c} h_{b}{}^{d} \nabla_c n_d,
\ee
where $n_d$ is the unit normalised future directed timelike normal and $h_{ab}$ is the induced metric on a  Cauchy surface near ${\cal I}^+$. For the Riemann tensor (both bulk and boundary) our conventions are same as Wald's textbook. de Sitter length $\ell$ has been set to unity. 

The counterterm charges are both conceptually and in form  different from ABK charges. It is natural to compare them. In the AdS context, the corresponding comparison was initiated by Ashtekar and Das~\cite{AshtekarDas}. This analysis was later completed by Hollands, Ishibashi, and Marolf~\cite{HIM}, who showed that the difference between the two charges is a ``constant'' offset, i.e.,  it does not depend on the particular asymptotic AdS spacetime under consideration. In this section following the work of Hollands et al, we show that in de Sitter context too, the difference is only a constant offset, expressible in terms of non-dynamical boundary data. This analysis is an extension of the technology we developed  for our covariant phase space analysis in the previous subsection. This analysis also shows that the trace of the counterterm stress tensor precisely matches with the trace-anomaly computed in~\cite{Nojiri:2001mf}, which is just the negative of the AdS result~\cite{Henningson:1998gx, Balasubramanian:1999re}.

 ABK charges \eqref{Charge_ABK} are defined on $\mathcal{I}^{+}$. Quantities $\tilde{E}_{ab}$, $\tilde u^b$, and $ \widetilde{dS}$ appearing in expression \eqref{Charge_ABK} refer to the unphysical metric. To compare expression \eqref{Charge_ABK} to counterterm  expression \eqref{charge_CT}, we start by writing $\tilde{E}_{ab}$, $\tilde u^b$ and  $ \widetilde{dS}$  in terms of physical variables. To this end we  need to use the conformal transformation $\tilde{g}_{ab}= \Omega^{2}{g}_{ab}$. Under this 
conformal transformation, $\tilde{n}^{c}=\Omega^{-1}{n}^{c}$, $\tilde{u}^{c}=\Omega^{-1} u^{c}  $, $
\widetilde{d S}=\Omega^{{d-2}}  dS$, and from definition of the electric part of the unphysical Weyl tensor \eqref{elec_weyl} we have
\be
(d-3) \tilde E_{ab} = \Omega^{3-d} \tilde{C}_{acbd}  \tilde{n}^{c} \tilde{n}^{d} 
= \Omega^{3-d} \Omega^{2}{C}_{acbd}  (\Omega^{-1} {n}^{c}) (\Omega^{-1}{n}^{d})
=\Omega^{3-d} C_{acbd}  n^{c} n^{d}.
\ee
Therefore,  in terms of physical variable expression \eqref{Charge_ABK} becomes, 
\be
H_{\xi}[C]=\lim_{C_\Omega \to C} \frac{1}{8 \pi G (d-3)}\int_{C_\O} (C_{acbd}  n^{c} n^{d}) \ {\xi}^{a} u^b dS. \label{ABK_our}
\ee

The ADM decomposition with respect to  $\O= \mbox{const}$ surfaces for the physical spacetime gives,
\be
(C_{acbd}  n^{c} n^{d}) =\cR_{ab}+{K}{K}_{ab}-{K}_{a}^{c}{K}_{cb}
-(d-2) h_{ab}. \label{electric_WEYL}
\ee
To compare \eqref{Charge_ABK} and \eqref{charge_CT} let us concentrate on the combination,
\bea
 \label{Comparison_Eq_MT}
8\pi G (d-3) {\tau}_{ab}-  (C_{acbd}  n^{c} n^{d})&=&
-\frac{1}{2}{h}_{ab}({K}_{mn}{K}^{mn}-{K}^{2})
-({K}{K}_{ab}-{K}_{a}^{c}{K}_{cb})\nn \\
& & +(d-3)({K}
{h}_{ab}-{K}_{ab})+\frac{(d-2)(d-3)}{2}{h}_{ab}.
\eea

As these two formalisms can only be compared at $\mathcal{I}^+$, it turns out to be most convenient  to express quantities appearing on the right hand side of \eqref{Comparison_Eq_MT}
in terms of the unphysical variables~\cite{AshtekarDas}. 
We decompose the \emph{unphysical} 
metric in ADM form,
\be
\tilde g_{ab}=- \tilde \eta_{a} \tilde \eta_{b}+\tilde h_{ab},
\ee 
where $\tilde \eta_a$ is unit timelike normal to $\Omega = \mbox{const}$ surfaces. For this discussion, we do not assume that $\tilde h_{ab}$ at $\Omega = 0$ is round metric on the sphere; at the end of the calculation we can specialise to that case. Moreover, we do \emph{not} necessarily work with $\O$ for which the unphysical metric take the Gaussian norm form \eqref{unphys_GN}. Let us define $\tilde n_{a} = \tilde \nabla_a \O$, and introduce, 
\be
 \tilde f:=\Omega^{-2}(1+\tilde \eta^{a}\tilde n_{a}),
 \label{def_f}
\ee
so that
\be
\tilde g_{ab}=-(1-\Omega^{2}\tilde f)^{-2} \tilde \nabla_{a}\Omega \tilde \nabla_{b}\Omega+\tilde h_{ab}(\Omega). \label{new_foliation_MT}
\ee
The function $ \tilde f$ has a smooth limit at $\mathcal{I}^+$~\cite{AshtekarDas}. The physical extrinsic curvature\footnote{Again, to avoid any possible confusion: $K_{ab} = - h_a{}^c h_b{}^d \nabla_c (\Omega^{-1} \tilde \eta_d).$} is related to the unphysical 
extrinsic curvature as,
\be
{K}_{ab}=\Omega^{-1}\tilde{K}_{ab}+\Omega^{-2}(\tilde{\eta} \cdot \tilde{n})\tilde{h}_{ab}. 
\ee
In terms of unphysical variables \eqref{Comparison_Eq_MT} becomes,
\be
8\pi G (d-3) {\tau}_{ab}-  (C_{acbd}  n^{c} n^{d}):=\Omega^{2} \tilde{\Delta}_{ab},
\ee
where for $d=5$,
\be
\Omega^{2} \tilde{\Delta}_{ab}
 =-\frac{1}{2}\big(\tilde{K}_{mn}\tilde{K}^{mn}-
\tilde{K}^2\big)\tilde{h}_{ab}-\tilde{K}\tilde{K}_{ab}+\tilde{K}_a ^c \tilde{K}_{cb}  + 2 \big(\tilde{K}\tilde{ h}_{ab}-\tilde{K}_{ab}\big)
(\Omega \tilde f)  +   3 \tilde{h}_{ab} (\O \tilde f)^2 .
 \label{CompCounterABK_MT}
\ee

A somewhat long calculation presented in appendix \ref{app:counter_ABK} shows that the unphysical extrinsic curvature $\tilde{K}_{ab}$ at $\mathcal{I}^+$ also satisfies,
\be
\Omega^{-1}\tilde K_{ab} = \frac{1}{(d-3)} \bigg(\tilde \cR_{ab} - \frac{1}{2(d-2)}\tilde \cR \tilde h_{ab}\bigg)-\tilde f \tilde h_{ab} \label{rescaled_K_MT},
\ee
where $\tilde \cR_{ab}$ and $\cR$ respectively  denote the Ricci tensor and the Ricci scalar of the induced metric at $\mathcal{I}^+$. From equation \eqref{rescaled_K_MT} it follows that $\tilde{\Delta}_{ab}$ has a smooth limit at $\mathcal{I}^+$. In five dimensions, the difference between counterterm charge and the ABK charge can be written as
\be
Q^{\rom{ct}}_\xi  [C] - H_\xi [C] = \frac{1}{16\pi G}\int_C \tilde \Delta_{ab} \tilde u^b \xi^a \widetilde {dS}.
\ee
Substituting \eqref{rescaled_K_MT} 
into equation \eqref{CompCounterABK_MT}, we get
\be
\tilde{\Delta}_{ab}=-\frac{1}{4}\bigg(\frac{2}{3} \tilde{\cR}\tilde{\cR}_{ab}-\frac{1}{4} \tilde{\cR}
^{2}\tilde{h}_{ab}-\tilde{\cR}_{a}^{c}\tilde{\cR}_{cb}+\frac{1}{2} \tilde{\cR}_{cd}\tilde{\cR}^{cd}
\tilde{h}_{ab}\bigg). \label{Delta}
\ee

In the above calculation we have not assumed that $\tilde h_{ab}$ at $\mathcal{I}^+$ is the round metric on the sphere. Therefore, the considerations of this subsection are slightly more general than of the previous subsections. As expected~\cite{Balasubramanian:1999re, HIM} the counterterm charges and the ABK charges differ. The difference, however, is a constant offset, which is determined by the curvature of the boundary metric alone. It does not depend on the specific asymptotically de Sitter solution. It can be evaluated on any asymptotically de Sitter solution given the boundary metric $\tilde h_{ab}$, in particular on pure de Sitter. The answer for pure de Sitter can be  compared with a calculation of the Casimir energy of the putative boundary theory as discussed in \cite{Klemm:2001ea}. We take five-dimensional de Sitter metric in static coordinates,
\be
ds^2 = -(1-r^2) dt^2  + (1-r^2)^{-1} dr^2 + r^2 d\sigma_{3}^2.
\ee
In these coordinates the canonical choice of the boundary metric is the standard metric on $\mathbb{R}~\times~\mathbb{S}^3$. We then have
\be
\tilde h_{ab} =  t_a t_b + \sigma_{ab},
\ee
where $\sigma_{ab}$ is the round metric on unit $\mathbb{S}^3$, $t^a = (\partial_t)^a$, and 
\be
\tilde \cR_{ab} = 2  \sigma_{ab}.
\ee
A small calculation then shows that, 
\be
Q^\rom{ct}_\xi  [C] - H_\xi [C] = - \frac{1}{16\pi G } \int_C \left(- \frac{3}{4} t_a t_b + \frac{1}{4} \sigma_{ab} \right)\tilde u^b \xi^a \widetilde {dS}. \label{comparison}
\ee
For dilatation $\xi^a =(\partial_t)^a$,
\be
Q^\rom{ct}_t  [C] - H_t [C]   = - \frac{3\pi}{32 G},
\ee
which precisely matches with the answer in~\cite{Klemm:2001ea}. It follows from equation \eqref{comparison} that the charges $Q^\rom{ct}_\xi  [C]$ are independent of the cross-section $C$ with the above choice of the boundary metric. This is because the  charges $H_\xi[C]$ have this property and the tensor $ \left(- \frac{3}{4} t_a t_b + \frac{1}{4} \sigma_{ab} \right)$ is traceless and covariantly conserved.
 
The difference \eqref{Delta} can also be compared with the trace anomaly computed in~\cite{Nojiri:2001mf}. The trace of $\tilde \Delta_{ab}$ is simply the trace of $\tau_{ab}$. We have
\be
\tau_{a}{}^{a} = - \frac{1}{64 \pi G} \left(\tilde \cR_{ab} \tilde \cR^{ab} - \frac{1}{3} \tilde \cR^2\right).
\ee
This answer is just the negative of the AdS result~\cite{Henningson:1998gx, Balasubramanian:1999re}. Thus the intuition that many of the de Sitter results with Dirichlet boundary conditions can be obtained from an analytic continuation of the corresponding AdS results continues to hold for this calculation too.

References \cite{Miskovic:2009bm, Jatkar:2014npa} showed that the analog of the ABK charges in AdS can be manifestly recovered from the addition of counterterms of an unusual sort. These counterterms involve extrinsic curvature of the boundary, unlike the ones used above. It is natural to speculate that a version of such results might also be valid in de Sitter. 

\section{de Sitter charges at spatial infinity}
\label{sec:MK_ABK}

In an interesting paper, Kelly and Marolf~\cite{Kelly:2012zc} pointed out that although there has been many discussions of de Sitter charges over the years, most of these discussions do not explicitly construct a phase space on which charges can be viewed as generators of the associated asymptotic symmetries. For the construction such as the one given above, induced metric at future infinity is held fixed and therefore the symplectic structure necessarily vanishes. In contrast, Kelly and Marolf imposed no such conditions. They proposed a definition of asymptotically de Sitter spacetimes (with non-compact Cauchy slices) with appropriate fall-off near spatial infinity $i^0$ without reference to $\cal{I}^+$.  Hence, their set-up is conceptually very different from the ABK set-up. Nonetheless, ABK commented that it is  likely that in the regime where the two approaches can be compared the two expressions for the charges agree.\footnote{Perhaps the intuition that such a comparison is possible comes from a related comparison at \emph{spatial} infinity in asymptotically \emph{flat} spacetimes~\cite{AM, MMV}.}  In this section we make this precise.

\subsection{Kelly-Marolf ADM charges}
\label{sec:MK_review}

For the convenience of the reader, we begin with a quick review of the Kelly-Marolf construction. We only focus on spatially flat Cauchy slices. Elements of the Kelly-Marolf phase space are globally hyperbolic solutions to Einstein equations with positive cosmological constant  that asymptote to Poincar\'e patch of de Sitter spacetime at spatial infinity. The metric on the Poincar\'e patch takes the following form in a standard set of coordinates,
\be
ds^2 = - dt^2 + e^{2t} \delta_{ab} dx^a dx^b,
\ee
where $a,b$ range over  $d-1$ cartesian coordinates (for simplicity we do not introduce separate notation for indices labelling Cartesian coordinates). For a general element of the phase space, introducing the time-function $t$ to define a spacelike foliation and choosing cartesian coordinates $x^a$ on each slice, the metric takes the general ADM form
\be
ds^2 = - N^2 dt^2 + h_{ab} (dx^a + N^a dt) (dx^b + N^bdt),
\ee 
with suitable boundary conditions for the spatial metric, extrinsic curvature, lapse and shift as $r = \sqrt{\delta_{ab} x^a x^b} \to \infty$. Proposed boundary condition for the spatial metric is
\be
\Delta h_{ab} = h_{ab} - \bar h_{ab} =  r^{-(d-1)}  h^{(d-1)}_{ab} + o(r^{-(d-1)}), \label{h_boud}
\ee
where $\bar h_{ab} = e^{2t} \delta_{ab}$ is the background spatial metric. The canonically conjugate momentum to $h_{ab}$ is 
\be
\pi^{ab} =  (K h^{ab} - K^{ab}), \label{Pi_def}
\ee
where $K_{ab}$ is the extrinsic curvature of constant $t$ slices. Boundary condition for $\pi^{ab}$ is
\be
\Delta \pi_{ab} = \pi_{ab}  - \bar \pi_{ab} = r^{-(d-2)}  \pi^{(d-2)}_{ab} + o(r^{-(d-2)}), \label{pi_boud}
\ee
where for the background,
\be
\bar \pi_{ab} = - (d-2) \bar h_{ab}. 
\ee
Finally, boundary conditions for the lapse and shift functions are,
\bea
N &=& 1 + r^{-(d-2)}  N^{(d-2)}+ o(r^{-(d-2)}) \label{N_boud} \\
N^a &=& r^{-(d-3)}  N^a_{(d-3)}+ o(r^{-(d-3)}). \label{Ni_boud}
\eea
As part of their boundary conditions, Kelly and Marolf also impose some further restrictions with regard to the even and odd nature of various coefficients on the asymptotic sphere $\mathbb{S}^{d-2}$. For our discussion we do not need those details explicitly, so, to keep things simple we do not write them. A key point being that these boundary conditions only restrict the behaviour in the limit $r\to \infty$, they do not restrict the induced metric at ${\cal I^+}$ directly at any finite $r$.

With these boundary conditions, asymptotic symmetries in $d > 3$ are simply the symmetries of de Sitter spacetime. 
For these asymptotic symmetries expression for charges is,
\be
Q^{\rom{KM}}_{\xi}[C]=\lim_{r \to \infty} \frac{1}{8 \pi G}\int_{(\partial \Sigma)_r}  \Delta' \pi_{ab} \ \xi^{a} u^{b} dS.
\label{charge_KM}
\ee
where
\be
\Delta' \pi^{ab} = \pi^{ab} + (d-2) h^{ab}, \label{DetlaPi}
\ee
$\Sigma$ is a slice that approaches $i^0$,  $ u^a$ is the unit normal to $(\partial \Sigma)_r$ in $\Sigma$.

Our aim is to show that these charges are same as the ABK charges. To compare we need to consider the overlap of Kelly-Marolf and ABK boundary conditions: 
a class of spacetimes which satisfy the above boundary conditions \eqref{h_boud}, \eqref{pi_boud},  \eqref{N_boud},  \eqref{Ni_boud} \emph{and} for which the induced metric at ${\cal I^+}$ is conformally flat. 

It is a priori  not clear for what classes of asymptotically de Sitter spacetimes these two boundary conditions are compatible. A full exploration of this question is beyond the scope of our work. We note that familiar examples, such as  Schwarzschild and Kerr-de Sitter metrics satisfy both these boundary conditions. Near spatial infinity, Kelly-Marolf boundary conditions require the spacetime to approach exact de Sitter spacetime and hence in that region the induced metric at ${\cal I}^+$ is fixed to be (conformally) flat metric. This reasoning was used to motivate the comparison between charges \eqref{charge_KM} and counter-term charges in~\cite{Kelly:2012zc}.

\subsection{Comparison with ABK  charges}
\label{sec:MK_ABK_comp}
In equation \eqref{ABK_our} we argued that ABK  charges can also be written as,
\be
H_{\xi}[C]=\lim_{C_\Omega \to C} \frac{1}{8 \pi G (d-3)}\int_{C_\O} (C_{acbd}  n^{c} n^{d}) \ {\xi}^{a} u^b dS.
\label{ABK_our2}
\ee
Consider a slice $\Sigma$ that approaches $i^0$. In order to facilitate comparison with \eqref{charge_KM} we can also write expression \eqref{ABK_our2} as 
\be
H_{\xi}[C]=\lim_{r \to \infty} \frac{1}{8 \pi G (d-3) }\int_{(\partial \Sigma)_r} (C_{acbd}  n^{c} n^{d}) \ {\xi}^{a} u^b dS.
\label{ABK_our3}
\ee
Motivated by our counterterm analysis, let us begin by looking at the combination,
\be
(d-3)\Delta' \pi_{ab} - (C_{acbd}  n^{c} n^{d}).
\ee
Using equations \eqref{DetlaPi} and \eqref{Pi_def} for $\Delta' \pi_{ab}$ and equation \eqref{electric_WEYL} for the electric part of the Weyl tensor, we have
\bea 
(d-3)\Delta' \pi_{ab} - (C_{acbd}  n^{c} n^{d}) &=& (d-3) (K h_{ab}-K_{ab})+ (d-2)(d-3) h_{ab}- \cR_{ab} \nn \\ 
 && - (KK_{ab}-K_{a}^{c}K_{cb})  +(d-2)  h_{ab}.
\eea
Upon replacing Ricci tensor with Einstein tensor, this expression can also be written as 
\bea
(d-3)\Delta' \pi_{ab} - (C_{acbd}  n^{c} n^{d}) &=& (d-3) (Kh_{ab}-K_{ab})- 
\left(\cG_{ab}+\frac{1}{2}\cR h_{ab}\right) \nn \\ & & -(KK_{ab}-K_{a}^{c}K_{cb}) + (d-2)^{2}h_{ab}
\eea
In appendix B ($k = 0$ case) of  reference~\cite{Kelly:2012zc}  it has been argued that $\cG_{ab}$ term can be neglected when we 
impose the above fall-off 
conditions.\footnote{More precisely what has been shown in reference~\cite{Kelly:2012zc} is that after expanding $h_{ab} = \bar h_{ab} + \Delta h_{ab}$ contribution of the $\cG_{ab}$ term is independent of $\Delta h_{ab}$ and $\Delta \pi_{ab}$, i.e., $\cG_{ab}$ term can at most yield an irrelevant shift of the charges that only depends on the background $\bar h_{ab}$.  For our discussion $\bar h_{ab}  = e^{2t} \delta_{ab}$, for which $\bar \cG_{ab} =0$ and hence the irrelevant shift of the charges also vanishes. Although the discussion in~\cite{Kelly:2012zc} is restricted to $d=4,5$, these comments are true in general dimension $d \ge 4$.} Hence, effectively, we have 
\bea
(d-3)\Delta' \pi_{ab}-(C_{acbd}  n^{c} n^{d})
&\simeq& (d-3) (Kh_{ab}-K_{ab})- \frac{1}{2}\cR h_{ab}  \nn \\  & & -(KK_{ab}-K_{a}^{c}K_{cb}) + (d-2)^{2}h_{ab},
\eea
which upon using contracted Gauss-Codazzi equation (to replace the Ricci scalar)  becomes, 
\bea
(d-3)\Delta' \pi_{ab}-(C_{acbd}  n^{c} n^{d})
&\simeq&
-\frac{1}{2}{h}_{ab}({K}_{cd}{K}^{cd}-{K}^{2})
-({K}{K}_{ab}-{K}_{a}^{c}{K}_{cb})\nn \\
& & +(d-3)({K}
{h}_{ab}-{K}_{ab})+\frac{(d-2)(d-3)}{2}{h}_{ab}.
\label{difference}
\eea
Note that the right hand side of equation \eqref{difference} is exactly the same expression as the right hand side of \eqref{Comparison_Eq_MT}.

Since Kelly-Marolf charges are defined at spatial infinity, we  are supposed to compute the above integrand near spatial infinity, i.e., as an expansion in inverse powers of $r$. To this end, we write
\begin{align}
& h_{ab} = \bar h_{ab} + \Delta h_{ab}, \label{expansion1}\\
& K_{ab} = \bar K_{ab} + \Delta K_{ab} = - \bar h_{ab} + \Delta K_{ab} , \label{expansion2} \\
& K = \bar K + \Delta K = - (d-1) + \Delta K,\label{expansion3} 
\end{align}
where as per the above boundary conditions in cartesian coordinates 
\begin{align}
&\Delta h_{ab} = {\cal O}(r^{-(d-1)}), &
&\Delta K_{ab} =  {\cal O}(r^{-(d-2)}), &
&\Delta K = {\cal O}(r^{-(d-2)}).&
\end{align}
Inserting expansions \eqref{expansion1}-\eqref{expansion3} in equation \eqref{difference} we observe that at the linear order in 
$\Delta h_{ab}$, $\Delta K_{ab}$, and $\Delta K$ all terms cancel out. The non-linear terms fall off much faster to contribute to surface integrals at spatial infinity. Thus, in the limit $r \to \infty$ the integrands that enters Kelly-Marolf definition of charges are the same as the integrands that enter ABK definition of charges. Therefore, the charges are the same at spatial infinity for the class of spacetimes that satisfy both  Kelly-Marolf and ABK boundary conditions.

In reference \cite{Bunster:2018yjr} another set of asymptotic conditions for de Sitter 
 spacetime near future infinity were explored. They also use the ADM canonical formalism, like Kelly and Marolf, but in a different foliation of spacetime that allows to cover all of ${\cal I}^+$.  We suspect that their boundary conditions and charges are also equivalent to those of ABK, but a detailed comparison is not attempted in this work. 
 
 We have also not attempted a direct comparison between ABK charges and Abbott-Deser charges \cite{Abbott:1981ff, Deser:2002jk}.

\section{Examples: Schwarzschild metric $d=4,5$}
\label{sec:examples}

In this section we analyse  Schwarzschild metric in detail. The aim is to express the (un)physical Schwarzschild metric in a form that manifests the fact it belongs to our covariant phase space, cf.~\eqref{FG_4d_MT}. From this form we want to extract the electric part of the Weyl tensor. 
For simplicity we focus on four- and five-dimensions in this section; generalisation to higher dimensions is straightforward.  Generalisation to rotating solutions is explored in appendix  \ref{app:examples}. We do not discuss computation of explicit charges for these solutions, as such calculations have already been discussed in the literature in a variety of contexts \cite{Hajian:2015xlp, Hajian:2016kxx, Deser:2005jf}.

Let us recall that the Schwarzschild-de sitter metric in coordinates that cover the region near the future infinity  is,
\be
ds^2 = -f dt^2 + f^{-1} dr^2 + r^2 d \sigma_{d-2}^2,
\ee
where $r$ is the timelike coordinate in the range $1 < r <\infty$  and
\be
f = 1 - r^2 - \frac{2m}{r^{d-3}}.
\ee
These coordinates are often called static coordinates, with future timelike infinity at $ r\to \infty$.
We have set de Sitter length $\ell = 1$. There are several conformal completions one can work with. We are interested in expressing the unphysical Schwarzschild-de Sitter metric in the Fefferman-Graham gauge~\eqref{FG_4d_MT}. To begin with we define 
\begin{align}
t & = \ln \tan (\chi/2), & r = \frac{1}{\Omega} \sin \chi,
\end{align}
and express the metric in terms of $\chi$ and $\Omega$.

In these new coordinates the physical metric takes the following form near the boundary $\Omega = 0$,
\bea
ds^2 &=& - \frac{d\Omega^2}{\Omega^2} \left[ 1 + \csc^2 \chi \ \Omega^2 - 2m \csc^3 \chi \ \Omega^3 + \mathcal{O}(\Omega^4)\right] \nn \\ 
& & +  \frac{2 d \chi d \Omega}{\Omega^2}  \left[   \cot \chi \ \Omega +  \cot \chi \csc^2 \chi \ \Omega^3 + \mathcal{O}(\Omega^4)\right] \nn  \\
& & + \ \frac{d \chi^2}{\Omega^2}  \left[ 1 -  ( \cot^2 \chi + \csc^2 \chi) \ \Omega^2 + 2m  \csc^3 \chi  \ \Omega^3 + \mathcal{O}(\Omega^4)\right]  \nn \\
& & + \  \frac{d \sigma^2}{\Omega^2} \sin^2 \chi,
\eea
where $d\sigma^2$ is the unit round metric on $
\mathbb{S}^2$. The unphysical metric $\Omega^2 ds^2$ is not in the Fefferman-Graham gauge;
though it is in the requisite form to \emph{zeroth} order. In order to show that the Schwarzschild-de Sitter solutions belongs to our phase space, we need to change this metric by changing $\Omega$ and $\chi$ coordinates such that we get rid of the cross term and make the coefficient of $d \Omega^2$ term to be $-\Omega^{-2}$ to requisite order. To this end, we introduce
new coordinates $\bar \Omega$ and $\bar \chi$, and correct them  from $\Omega$ and $\chi$ via 
 four functions $h_1(\bar \chi), h_2(\bar \chi), k_1(\bar \chi), k_2(\bar \chi)$ in power series in $\bar \Omega$:
 \bea
\Omega &=& \bar \Omega + h_1(\bar \chi) \bar \Omega^3 + h_2(\bar \chi) \bar \Omega^4 + \ldots \\
\chi &=& \bar \chi + k_1(\bar \chi) \bar \Omega^2 + k_2(\bar \chi) \bar \Omega^4 + \ldots~.
\eea
The following choice uniquely gives the unphysical metric in the desired  form
\begin{align}
k_1(\bar \chi) &= - \frac{1}{2} \cot \bar \chi, &
h_1(\bar \chi) &=  -\frac{1}{4} (\csc^2 \bar \chi + \cot^2 \bar \chi), &\\
k_2(\bar \chi) &= -\frac{1}{8} \cot \bar \chi \csc^2 \bar \chi,&
h_2(\bar \chi) &= \frac{1}{3} m \csc^3 \bar \chi.&
\end{align}
These functions serve the following purposes:  $k_1$ ensures that $\tilde g_{\bar\Omega \bar \chi}$ vanishes at order $\bar \Omega$; $h_1$ ensures that $\tilde g_{\bar\Omega\bar\Omega}$ is $-1$ at order  $\bar\Omega^2$; having fixed $k_1$ and $h_1$, $k_2$ ensures that $\tilde g_{\bar\Omega \bar \chi}$ vanishes at orders $\bar \Omega^3$; and finally, $h_2$ ensures that $\tilde g_{\bar\Omega\bar\Omega}$ is $-1$  at order $\bar \Omega^3$.

After these changes, the unphysical metric becomes
\bea
d\tilde s^2 &=& \bar \Omega^2 ds^2 
\ =  \ - d \bar \Omega^2 + ( d \bar \chi^2 +  \sin^2 \bar \chi \ d \sigma^2) \left( 1 + \frac{1}{2}\bar \Omega^2 \right) \nn \\
&& \quad \quad \quad \quad \quad + \ \frac{4}{3} m \csc^3 \bar \chi \ \bar \Omega^3 \ d \bar \chi^2 - \frac{2}{3} m \csc \bar \chi \ \bar \Omega^3 \ d\sigma^2+ \mathcal{O}(\bar \Omega^4).
\eea
From this form of the metric we can simply read off the electric part of the Weyl tensor, cf \eqref{FG_4d_MT}. We find
\begin{align}
\tilde E_{\bar \chi \bar \chi} &= - 2m \csc^3  \bar \chi, &
\tilde E_{\theta_i \theta_i} &= m \csc  \bar \chi \ g_{\theta_i \theta_i} ,& 
\end{align}
where $g_{\theta_i \theta_i}$ are components of the round metric on $\mathbb{S}^{2}$.
We can compare this tensor with the expressions given in~\cite{ABKI} by ABK. ABK work with the boundary metric to be natural metric on $\mathbb{R} \times \mathbb{S}^2$,
\be
ds^2_{\mathbb{R} \times \mathbb{S}^2} = dt^2 + (d\theta^2 + \sin^2 \theta d \phi^2).
\ee
The change of coordinates that takes from the natural metric on $\mathbb{R} \times \mathbb{S}^2$ to the unit round metric on  $\mathbb{S}^3$ is 
$t  = \ln \tan (\bar \chi/2)$, together with the conformal factor $\omega = \sin \bar \chi$:
\be
ds^2_{\mathbb{S}^3} = \sin^2 \bar \chi \left[\frac{d \bar \chi^2}{\sin^2 \bar \chi}  +  (d\theta^2 + \sin^2 \theta d \phi^2)\right].
\ee
Moreover, ABK write
\begin{align}
\tilde E_{tt}^\rom{ABK} &= - 2m,& 
\tilde E_{\theta_i \theta_i}^\rom{ABK} &=  m g_{\theta_i \theta_i}, &
\end{align}
These expressions are perfectly consistent with tensor transformation of a conserved traceless tensor under conformal and coordinate transformation (see e.g., Wald's textbook~\cite{Wald} discussion around equation (D.20))
\be
\tilde E_{ab} = \omega^{-(d-1)+2} \tilde E_{ab}^\rom{ABK}\bigg{|}_{d=4} = \csc \chi \ \tilde E_{ab}^\rom{ABK}.
\ee

To summarise: with this computation we have completed a full circle. We have shown that the Schwarzschild-de Sitter solution belongs to our  phase space. We extracted the electric part of the Weyl tensor by expressing Schwarzschild-de Sitter solution in Fefferman-Graham gauge. Since electric part of the Weyl tensor is traceless and conserved on the boundary manifold, we can relate our extracted electric part of the Weyl tensor to the expressions given by ABK via \emph{boundary} conformal and coordinate transformation.  The expressions given by ABK are obtained in a logically different way: they obtained it via directly computing the relevant components of the  four-dimensional unphysical Weyl tensor. The fact that we are able to relate these expressions in a expected way provides a non-trivial consistency check on all our computations. 

 In appendix \ref{app:examples}  four-dimensional Kerr-de Sitter metric is expressed in Fefferman-Graham gauge and a similar computation relating different ways of obtaining electric part of the Weyl tensor is performed.

Let us now indicate how the above computation works for $d=5$. The over-all logic remains almost exactly the same. The physical metric takes the form near the boundary $\Omega = 0$,
\bea
ds^2 &=& - \frac{d\Omega^2}{\Omega^2} \left[ 1 + \csc^2 \chi \ \Omega^2 - \left( 2m  -1  \right)\csc^4 \chi \ \Omega^4+ \mathcal{O}(\Omega^6)\right] \nn \\ 
& & +  \frac{2 d \chi d \Omega}{\Omega^2}  \left[   \cot \chi \ \Omega +  \cot \chi \csc^2 \chi \ \Omega^3 -\left( (2m-1)\cot \chi \csc^4 \chi \right) \Omega^5+ \mathcal{O}(\Omega^7)\right] \nn  \\
& & + \ \frac{d \chi^2}{\Omega^2}  \left[ 1 -  \left( \cot^2 \chi + \csc^2 \chi \right) \ \Omega^2 +\left( -\cot^2 \chi \ \csc^2 \chi+2m \csc^4 \chi\right) \ \Omega^4 + \mathcal{O}(\Omega^6)\right]  \nn \\
& & + \  \frac{d \sigma^2}{\Omega^2} \sin^2 \chi,
\eea
where $d \sigma^2$ is now the round metric on unit $\mathbb{S}^3$. This metric is not in the Fefferman-Graham gauge. In order to show that the 5d Schwarzschild-de Sitter solution belongs to our  phase space, we need to change this metric by changing $\Omega$ and $\chi$ as above:
\bea
\Omega &=& \bar \Omega + h_1(\bar \chi) \bar \Omega^3 + h_2(\bar \chi) \bar \Omega^5 + \ldots \\
\chi &=& \bar \chi + k_1(\bar \chi) \bar \Omega^2 + k_2(\bar \chi) \bar \Omega^4 + \ldots~.
\eea
The following choice uniquely gives the metric in the desired  form
\begin{align}
k_1(\bar \chi) &= - \frac{1}{2} \cot \bar \chi, &
h_1(\bar \chi) &=  -\frac{1}{4} (\csc^2 \bar \chi + \cot^2 \bar \chi), & \\
k_2(\bar \chi) &= -\frac{1}{8} \cot \bar \chi \csc^2 \bar \chi, &
h_2(\bar \chi) &= \frac{1}{8} \left(\frac{1}{2}+\csc^2 \bar \chi+ (2m-1) \csc^4 \bar \chi\right) , &
\end{align}
These functions serve the following purposes:  $k_1$ ensures that $\tilde g_{\bar\Omega \bar \chi}$ vanishes at order $\bar \Omega$; $h_1$ ensures that $\tilde g_{\bar\Omega\bar\Omega}$ is $-1$ at order  $\bar\Omega^2$; having fixed $k_1$ and $h_1$, $k_2$ ensures that $\tilde g_{\bar\Omega \bar \chi}$ vanishes at orders $\bar \Omega^3$; and finally, $h_2$ ensures that $\tilde g_{\bar\Omega\bar\Omega}$ is $-1$  at order $\bar \Omega^4$.

The resulting  unphysical metric is
\bea
d\tilde s^2
&=& - d \bar \Omega^2 + ( d \bar \chi^2 +  \sin^2 \bar \chi \ d \sigma^2) \left( 1 + \frac{1}{2}\bar \Omega^2 + \frac{1}{16}\bar \Omega^4\right) \nn \\
&& - \ \frac{1}{2} \bar \Omega^4 \left(-3m \csc^4 \bar \chi  \ d \bar \chi^2 + m \csc^2 \bar \chi  d\sigma^2 \right)+ \mathcal{O}(\bar \Omega^5).
\eea
From this form of the metric we can simply read off the electric part of the Weyl tensor.
We find
\begin{align}
\tilde E_{\bar \chi \bar \chi} &= -3m \csc^4 \bar \chi,  &
\tilde E_{\theta_i \theta_i} &=  m \csc^2 \bar \chi \ g_{\theta_i \theta_i} ,& 
\end{align}
where $g_{\theta_i \theta_i}$ are components of metric on the round $\mathbb{S}^{3}$.
The analog of ABK expressions for five-dimensional  Schwarzschild-de Sitter solution are
\begin{align}
\tilde E_{tt}^\rom{ABK} &= - 3m,& 
\tilde E_{\theta_i \theta_i}^\rom{ABK} &=  m g_{\theta_i \theta_i}, &
\end{align}

Once again, one can easily check that these expressions are perfectly consistent with each other.
The change of coordinates that takes from natural metric on $\mathbb{R} \times \mathbb{S}^3$ to unit metric on  $\mathbb{S}^4$ is 
$t  = \ln \tan (\bar \chi/2)$, together with the conformal factor $\omega = \sin \bar \chi$. Under this transformation,
\be
\tilde E_{ab} = \omega^{-(d-1)+2} \tilde E_{ab}^\rom{ABK}\bigg{|}_{d=5} = \csc^2 \chi \ \tilde E_{ab}^\rom{ABK}.
\ee

 In appendix \ref{app:examples} five-dimensional Myers-Perry-de Sitter metric is expressed in Fefferman-Graham gauge and a similar consistency check is performed.


\subsection*{Acknowledgements}
We thank Abhay Ashtekar, Bidisha Chakrabarty, Aniket Khairnar, Alok Laddha, Donald Marolf, and K Narayan for discussions on various aspects of this project. Our work is supported in part by Max-Planck Partner Group ``Quantum Black Holes'' between CMI Chennai and AEI Potsdam. 

\appendix


\section{Details on asymptotic expansion} 

\label{App_asym}

\subsection*{ADM decompostion} 
In this appendix we present a discussion of asymptotic expansion, in particular a derivation of the evolution equations \eqref{FG1}-\eqref{FG3} following~\cite{HIM}. This analysis was first carried out in~\cite{Jaeger}. 
Einstein equation with a positive cosmological constant is,
\be
R_{ab} - \half R g_{ab} + \La g_{ab}=0,
\ee
with 
\be
\La = \frac{(d-1)(d-2)}{2\ell^2}.
\ee
For the most part, we will be working with the unphysical metric is
\be
\tilde{g}_{ab} = \O^2 g_{ab}.
\ee 
The Ricci tensor $\tilde{R}_{ab}$   of the unphysical metric obtained by conformal transformation and upon using the equations of motion for the physical metric takes the form,
\be
\tilde{R}_{ac} = -(d-2)\O^{-1} \tilde{\nabla}_a \tilde{\nabla}_c \O - \O^{-1} \tilde{g}_{ac}  
\tilde{\Box} \O + (d-1) \O^{-2} \tilde{g}_{ac} \left( \tilde{\nabla}_d \O \tilde{\nabla}^d \O 
+ 1/\ell^2 \right). \label{tildeRab}
\ee
Taking the trace of \eqref{tildeRab} we get the Ricci scalar of the unphysical metric,
\be
\tilde{R} = -2(d-1)\O^{-1} \tilde{\Box} \O + d(d-1) \O^{-2} \left(\tilde{\nabla}_c \O \tilde{\nabla}^c \O + 1/ \ell^2 \right).
\ee
Defining, 
\be
\tilde{S}_{ab} := \frac{2}{d-2} \tilde{R}_{ab} - \frac{1}{(d-1)(d-2)} \tilde{R}\tilde{g}_{ab},
\ee
Einstein equations can be rewritten as,
\be
\tilde{S}_{ab} + 2 \O^{-1} \tilde{\nabla}_a\tilde{\nabla}_b \O - \O^{-2} \tilde{g}_{ab} \left(\tilde{\nabla}^c \O \tilde{\nabla}_c \O + 1/\ell^2 \right) = 0.
\ee
Multiplying the above equation by $\O^2$ and evaluating it at $\mathcal{I}^{+}$, we get 
\be
\tilde \nabla_c \O \tilde \nabla^c \O = - 1/ \ell^2 \label{grad_Omega}\qquad \mbox{at} \qquad \mathcal{I}^{+}.
\ee 
For simplicity, we henceforth work with $\ell=1$. Although  equation \eqref{grad_Omega} holds only at $\cI^+$, it can be made to hold in a neighbourhood of $\cI^+$~\cite{HIM}, i.e., Gaussian normal coordinates near $\cI^+$ can be chosen so that the unphysical metric takes the form
\be
\widetilde{(ds)}^2 := \tilde g_{ab} dx^a dx^b = - d\Omega^2  + \tilde h_{ab}(\Omega) dx^a dx^b.
\ee
Einstein equation then reads, 
\be
\tilde{S}_{ab} + 2 \O^{-1} \tilde{\nabla}_a\tilde{n}_b = 0, \label{einstein}
\ee
where $\tilde{n}_a = \tilde{\nabla}_a \Omega $ is the  timelike  unit normal to  $\O$ = constant hypersurfaces. 

Now we perform the ADM decomposition and rewrite \eqref{einstein} into evolution and constraint equations. To write the constraint equations in terms of the unphysical metric, we need the unphysical Einstein tensor $\tilde{G}_{ab}$ which is given by,
\be
\tilde{G}_{ab} = (d-2) \O^{-1} ( \tilde{K}_{ab} - \tilde{g}_{ab}\tilde{K} ), \label{unphys_eins}
\ee
where $\tilde{K}_{ab} = - \tilde{\nabla}_a \tilde{n}_b = - \tilde{\nabla}_a\tilde{\nabla}_b \O $. Using the Gauss-Codazzi equations together with \eqref{unphys_eins}, we get our two constraint equations,
\bea
 \tilde{\cR} + \tilde{K}^2 - \tilde{K}_{ab} \tilde{K}^{ab} &=& 2 (d-2) \O^{-1} \tilde{K} \\
\tilde{D}_c \tilde{K}^c_{\ a} - \tilde{D}_a \tilde{K} &=& 0,
\eea
and two evolution equations, 
\bea 
\pounds_n \tilde{h}_{ab} &=& -2\tilde{K}_{ab}, \\
\pounds_n \tilde{K}_a^{\ b} &= & \tilde{\cR}_a^{\ b} + \tilde{K} \tilde{K}_a^{\ b} - \O^{-1} \tilde{K} \tilde{h}_a^{\ b} - (d-2) \O^{-1} \tilde{K}_a^{\ b} \label{evo_K},
\eea 
where $\tilde \cR_{ab}$ and $\tilde \cR$ respectively  denote the Ricci tensor and the Ricci scalar of the metric $\tilde h_{ab}(\Omega)$ and $\tilde{D}_c$ is the unique torsionless derivative compatible with $\tilde h_{ab}(\Omega)$.

It is convenient to write \eqref{evo_K}  in terms of its trace and trace-free parts separately.  Defining, 
\be
\tilde{p}_a{}^{b} = \tilde{K}_a{}^{b} - \frac{\tilde h_a{}^b}{(d-1)} \tilde{K},
\ee
and using \be \pounds_n \equiv  - \frac{d}{d\O}, \ee we have
\bea
&& \frac{d}{d \O} \ \tilde{p}_a^{\ b} ~=~ - \left[ \tilde{\cR}_a^{\ b} - \frac{\tilde h_a{}^b}{(d-1)} \tilde{\cR} \right] - \tilde{K} \tilde{p}_a{}^{b} + (d-2) \O^{-1} \tilde{P}_a{}^{b} , \label{evo1}\\
&& \frac{d}{d \O} \ \tilde{K}_{\ }  ~=~ - \tilde{R} - \tilde{K}^2 + (2d-3)\O^{-1} \tilde{K}, \label{evo2}\\
&& \frac{d}{d \O} \ \tilde{h}_{ab} ~=~   2 \tilde{h}_{bc}\tilde{K}^c_{\ a} \label{evo3}.
\eea
Asymptotic expansion of these equations is obtained upon substituting a Taylor expansion for the metric in powers of $\O$. Inserting 
\eqref{h_Taylor}-\eqref{R_Taylor} in equations \eqref{evo1} to \eqref{evo3} we get \eqref{FG1}-\eqref{FG3}.

\subsection*{Asymptotic form}
As discussed in section \ref{asym}  recursions  \eqref{FG1}-\eqref{FG3} continue to all orders if $(\tilde p_{a}{}^b)_{d-2}$ is known.  Now we show that this information is contained in the electric part of the Weyl tensor.  Using Gauss-Codazzi equations, it can be shown that
\be
\tilde{C}_{abcd} \tilde{n}^b \tilde{n}^d = \pounds_n \tilde{K}_{ac} + \tilde{K}_a^{\ b}\tilde{K}_{bc} + \O^{-1} \tilde{K}_{ac}.
\ee
Taylor expanding the Weyl tensor in powers of $\O$, we get
\be
\left( \tilde{C}_{abcd} \tilde{n}^b \tilde{n}^d \right)_{(j-1)} = - (j-1) \left( \tilde{K}_{ac} \right)_j + \sum_{m=0}^{j-1} \left(\tilde{K}_a^{\ b} \right)_m \left(\tilde{K}_{bc} \right)_{j-1-m}
\ee
For $j=d-2$, this equation becomes,
\be
\frac{1}{d-3} \left( \tilde{C}_{abcd} \tilde{n}^b \tilde{n}^d \right)_{(d-3)} = - \left( \tilde{K}_{ac} \right)_{d-2} + \sum_{m=0}^{d-3} \frac{1}{d-3} \left(\tilde{K}_a^{\ b} \right)_m \left(\tilde{K}_{bc} \right)_{d-3-m}.
\ee
We define the electric part of the Weyl tensor as,
\be
\tilde{E}_{ac} = \frac{1}{d-3} \O^{3-d} \left( \tilde{C}_{abcd} \tilde{n}^b \tilde{n}^d \right),
\ee
from which it follows that,
\be
(\tilde{E}_{ac})_0 = \frac{1}{d-3}  \left( \tilde{C}_{abcd} \tilde{n}^b \tilde{n}^d \right)_{d-3} = - \left( \tilde{K}_{ac} \right)_{d-2} + \sum_{m=0}^{d-3} \frac{1}{d-3} \left(\tilde{K}_a^{\ b} \right)_m \left(\tilde{K}_{bc} \right)_{d-3-m}.
\label{FGWeyl}
\ee
Let us analyse these equations in different dimensions.

\subsubsection*{Four dimensions} Let us start with an analysis in $d=4$. From \eqref{FGWeyl} we have
\be
(\tilde E_{ab})_0  = (\tilde C_{abcd} \tilde n^c \tilde n^d)_{1} = - (\tilde K_{ab})_{2},
\ee
which from the recursions \eqref{FG1}-\eqref{FG3} implies
\be
(\tilde h_{ab})_{3} =  \frac{2}{3} (K_{ab})_2 =-   \frac{2}{3}  (\tilde E_{ab})_0,
\ee
and hence \eqref{FG_4d_MT}.

\subsubsection*{Five dimensions} 
Let us now write equation \eqref{FG1} for $d=5$. Setting $j=1$ we have
\be
 (\tilde p_{a}{}^b)_1 = \frac{1}{2} (\tilde{\cal R}_a{}^{b} )_{0} - \frac{1}{8}(\tilde {\cal  R} )_{0} \delta_a{}^{b}  =0
 \ee
 since $({\cal R}_{ab})_0 = 3 (h_{ab})_0$ for the four-sphere (recall that for $n$-sphere, ${\cal R}_{ab} = (n-1) h_{ab}$) and $(K_{ab})_0 = 0$. From \eqref{FG2} we get
\be
(\tilde K)_1 = \frac{1}{6}(\tilde{ \cal  R} )_{0} = 2.
\ee
Using these inputs we have
\be
(\tilde K_{ab} )_1 = (\tilde p_{a}{}^{c})_1 (\tilde h_{cb})_0 + \frac{1}{4} (\tilde h_{ab})_0 (\tilde K)_1  = \frac{1}{2} (\tilde h_{ab})_0.
\ee
Proceeding further, putting $d=5$ in equation \eqref{FGWeyl} we get
\be
(\tilde E_{ab})_0 = \frac{1}{2} (\tilde C_{abcd} \tilde n^c \tilde n^d)_{2} = -(\tilde K_{ab})_{3} + \frac{1}{2}(\tilde K_{ac})_1 (\tilde K_b{}^c)_{1} =- (\tilde K_{ab})_{3} +\frac{1}{8} (\tilde h_{ab})_0
\ee
and
\be
(\tilde h_{ab})_{4} =  \frac{1}{2} (K_{ab})_3 =  - \frac{1}{2} (\tilde E_{ab})_0 + \frac{1}{16} (\tilde h_{ab})_0.
\ee

\subsubsection*{Six and higher dimensions} For six and higher dimensions, the analysis becomes simpler. We have,
\be
(\tilde h_{ab})_{d-1} =  \frac{2}{d-1} (\tilde K_{ab})_{d-2},
\ee
and
\be
(\tilde E_{ab})_0 = \frac{1}{d-3} (\tilde C_{abcd} \tilde n^c \tilde n^d)_{d-3} = - (\tilde K_{ab})_{d-2} + \frac{1}{d-3}\sum_{m=0}^{d-3}(\tilde K_{ac})_m (\tilde K_b{}^c)_{d-3-m}.
\ee
For $d\ge6$, the sum in the second term vanishes. The reason is as follows. For pure de Sitter metric  $(\tilde h_{ab})_{d-1} =0$ and also $\tilde C_{abcd}=0$. Therefore, the sum vanishes for pure de Sitter. For general asymptotically de Sitter metric,  to evaluate the sum we only need to know $(\tilde K_{ab})_j$ for $j \le (d-3)$, which are all determined from the initial data $(h_{ab})_0$ through the equations \eqref{FG1}--\eqref{FG3}. The value of sum is therefore identical to its value for pure de Sitter, which vanishes. We conclude that
\be
(\tilde h_{ab})_{d-1}  = - \frac{2}{d-1} (\tilde E_{ab})_0.
\ee

\subsection*{Presymplectic potential on $\mathcal{I^{+}}$}
In order to investigate the behavior of the presymplectic potential $\theta(g, \delta g)$ at $\mathcal{I^+}$, we begin by looking at the asymptotic behavior of the following quantities,
\bea
\delta g_{ab}&=&-\frac{2}{d-1} \Omega^{d-3}(\delta {\tilde{E}}_{ab}) +\mathcal{O}
(\Omega^{d-2}) \label{gamma_app},\\
\delta \tilde{E}_{ab} \ \tilde{g}^{ab}&=&-\tilde{E}_{ab} \ 
\delta \tilde{g}^{ab}=\tilde{E}^{ab} \ \delta \tilde{g}_{ab}= \mathcal{O}(\Omega^{d-1}),\label{traceE} \\
\delta\tilde{n}^{a}&=&\delta \tilde{g}^{ac} \ \tilde{\nabla}_{c}\Omega
=\frac{2}{d-1} \Omega^{d-1} \delta \tilde{E}^{ab} \ \tilde{n}_{b}+
\mathcal{O}(\Omega^d)=\mathcal{O}(\Omega^d)\\
\tilde{n}^{a}\delta\tilde{E}_{ab}&=&-\tilde{E}_{ab}\delta \tilde{n}^{a}= 
\mathcal{O}(\Omega^d)
\eea
where we have used $\tilde{E}^{ab}  \tilde{g}_{ab} = 0$ and $\tilde{E}^{ab} \ \tilde{n}_{b}=0$.  Similarly, 
\bea
\delta \tilde{\epsilon}_{ab\cdots c}
&=&\frac{1}{2} \tilde g {}^{ef} \delta \tilde g_{ef} \ \tilde{\epsilon}_{ab\cdots c}  \\ 
&=& -\frac{1}{d-1} \Omega^{d-1}\delta {\tilde{E}}_{d}{}^{d} \tilde{\epsilon}_{ab\cdots c}+\mathcal{O}(\Omega^d) \\ 
&=& \mathcal{O}(\Omega^{d}),  \qquad \mbox{for~}  d > 2,
\eea
where we have used equations \eqref{gamma_app} and \eqref{traceE}. Similar manipulations give, 
\bea
\tilde{g}^{ac} \nabla_{a} \delta\tilde{E}_{bc}&=&\tilde{g}^{ac} \tilde{\nabla}_{a} \delta
\tilde{E}
_{bc}+\frac{1}{\Omega}\big(\tilde{n}_{b} \ \delta\tilde{E}_{ac}\tilde{g}^{ac}+(2-d) \tilde{n}
^{a} \ \delta\tilde{E}_{ab}\big)\\
&=& \mathcal{O}(\O^{0})+\frac{1}{\Omega}\big(\tilde{n}_{b} \ \delta\tilde{E}_{ac}\tilde{g}
^{ac}+(2-d) \tilde{n}^{a} \ \delta\tilde{E}_{ab}\big)\\
&=& \mathcal{O}(\O^{0}),  \qquad \mbox{for~}  d > 2.
\eea

With this preparation, we can now  investigate the asymptotic behavior of the presymplectic potential 
\eqref{PreSymPot},
\bea
16 \pi G \ \theta_{a_1 \ldots a_{d-1}} &=&  \epsilon_{c a_1 \ldots 
a_{d-1}} \ 
g^{ce}g^{bd}(\nabla_{d}\delta{g}_{be}-\nabla_{e}\delta{g}_{bd})\\
&=&\O^{4-d} \ \tilde{\epsilon}_{c a_1 \ldots a_{d-1}} \tilde g^{ce} \tilde g^{bd}
(\nabla_{d}\delta{g}_{be}-\nabla_{e}\delta{g}_{bd})\\
&=& 2\bigg(\frac{d-3}{d-1}\bigg)\tilde{\epsilon}_{c a_1 \ldots a_{d-1}}
\big(\tilde{g}^{ce}\tilde{n}^{b} \delta \tilde{E}_{be}-\tilde{n}^{c}\tilde{g}^{bd}\delta\tilde{E}
_{bd}\big)\\
&&-\frac{2}{d-1} \ \tilde{\epsilon}_{c a_1 \ldots a_{d-1}} \O  \ \tilde{g}^{ce}\tilde{g}^{bd}
\big(\nabla_{d}\delta \tilde{E}_{be}-\nabla_{e}\delta \tilde{E}_{bd}\big)\\
&=&\mathcal{O}(\O^{d-1}) -\frac{2}{d-1} \ \tilde{\epsilon}_{c a_1 \ldots a_{d-1}} \O  \ 
\tilde{g}^{ce}\tilde{g}^{bd}
\big(\nabla_{d}\delta \tilde{E}_{be}-\nabla_{e}\delta \tilde{E}_{bd}\big)\\
&=&\mathcal{O}(\O^{d-1}) -\frac{2}{d-1} \ \tilde{\epsilon}_{c a_1 \ldots a_{d-1}} \O  \ 
\bigg(\mathcal{O}(\O^0)-\tilde{g}^{ce}\tilde{g}^{bd}\nabla_{e}\delta \tilde{E}_{bd}\bigg)\\
&=&\mathcal{O}(\O)-\frac{2}{d-1} \ \tilde{\epsilon}_{c a_1 \ldots a_{d-1}} \bigg(\tilde{g}
^{ce} \underbrace{\tilde{n}^{d}\delta \tilde{E}_{ed}}_{\mathcal{O}(\O^d)}+2 \tilde{n}^{c}
\underbrace{\tilde{g}^{bd}\delta \tilde{E}_{bd}}_{\mathcal{O}(\O^{d-1})}+
\tilde{g}^{ce}\underbrace{\tilde{n}^{b}\delta \tilde{E}_{eb}}_{\mathcal{O}(\O^d)}\bigg)\\
&=& \mathcal{O}(\O)
\eea
Hence,
\be
\theta(g, \delta g) \  \Big{|}_{\mathcal{I}^+}=0.
\ee

\section{Details on comparison between counterterm and ABK charges}
\label{app:counter_ABK}

In this appendix we present a derivation of equation \eqref{rescaled_K_MT} relating $\tilde K_{ab}$ and $\tilde \cR_{ab}$. This equation in AdS context was written in~\cite{HIM}, though no details were given. Here we fill in those details.  Let us start with definition of $\tilde K_{ab}$ for $\O = \mbox{const}$ slices in foliation \eqref{new_foliation_MT}, 
\bea
\tilde K_{ab}&:=&-\tilde h^{c}_{a} \tilde h^{d}_{b} \tilde \nabla_{c} \tilde \eta_{d} \label{tilde_K1}\\
      &=& \tilde h^{c}_{a} \tilde h^{d}_{b} \  \tilde \nabla_{c}\big\{ (\tilde \eta \cdot \tilde n)^{-1}\tilde n_{d}\big\}
      \label{tilde_K2} \\
         &=&\tilde  h^{c}_{a} \tilde h^{d}_{b} (\tilde \eta \cdot \tilde n)^{-1}  \ \tilde \nabla_{c} \tilde n_{d}\label{tilde_K3}
\eea
where in going from \eqref{tilde_K1} to \eqref{tilde_K2} we have used the fact that unit normal $\tilde \eta_a$ is proportional to $\tilde n_a = \tilde \nabla_a \O$ in foliation \eqref{new_foliation_MT}. Since both these vectors are timelike and future directed, the proportionality factor is  with a minus sign, 
\be
\tilde \eta_{a} =  - (\tilde \eta \cdot \tilde n)^{-1}\tilde n_{a}.
\ee
In going from  \eqref{tilde_K2} to \eqref{tilde_K3} we have used the fact that $h^{d}_{b}\tilde n_{d} = 0$. Expanding out  $\tilde  h^{c}_{a} \tilde h^{d}_{b}$ factors in equation \eqref{tilde_K3} we get, 
\bea
(\tilde \eta \cdot \tilde n)\tilde K_{ab}&=&  \tilde h^{c}_{a} \tilde h^{d}_{b}  \ \tilde \nabla_{c} \tilde n_{d}\\
&=&(\delta^{c}_{a}+\tilde \eta^{c} \tilde \eta_{a})(\delta^{d}_{b}+\tilde \eta^{d}\tilde \eta_{b})           
 \ \tilde \nabla_{c} \tilde n_{d}\\
 &=&(\delta^{c}_{a}+\tilde \eta^{c} \tilde \eta_{a}) \ ( \tilde \nabla_{c} \tilde n_{b}+\tilde \eta^{d}\tilde \eta_{b}         
                                                                                       \tilde  \nabla_{c} \tilde n_{d})                                                                     
\eea

Now let us concentrate on the term $\tilde \eta^{d}\tilde \eta_{b} \tilde \nabla_{c} \tilde n_{d}$. Using definition \eqref{def_f} we have, 
\bea
\tilde \eta^{d} \tilde \eta_{b} \tilde \nabla_{c} \tilde n_{d}=\tilde \eta^{d}\tilde \eta_{b} \tilde \nabla_{c}\big\{ (1-\Omega^{2}\tilde f) 
\tilde \eta_{d}\big\}
=\tilde \eta_{b}\tilde \nabla_{c}(\Omega^{2}\tilde f),
\eea       
where have used $\tilde \eta^{d}\tilde \eta_{d}=-1$ and $\tilde \eta^{d} \nabla_{c}\tilde \eta_{d}=0$.
Therefore,
\bea
(\tilde \eta \cdot \tilde n)\tilde K_{ab}&=&(\delta^{c}_{a}+\tilde \eta^{c} \tilde \eta_{a}) \ \big(\tilde \nabla_{c} \tilde n_{b}
                  + \tilde \eta_{b} \tilde \nabla_{c}(\Omega^{2}\tilde f)\big)\\                                                                           
        &=&\big( \tilde \nabla_{a}\tilde n_{b}+\tilde \eta_{b}\tilde \nabla_{a}(\Omega^{2}\tilde f)+\tilde \eta^{c} \tilde \eta_{a}
                        \tilde    \nabla_{c}\tilde n_{b}+\tilde \eta^{c}\tilde \eta_{a}\tilde \eta_{b} \tilde \nabla_{c}(\Omega^{2}\tilde f)\big)\\
        &=& \big( \tilde \nabla_{a}\tilde n_{b}+\tilde \eta_{b}\tilde \nabla_{a}(\Omega^{2}\tilde f)+\tilde \eta^{c}\tilde \eta_{a}
                           \tilde  \nabla_{b} \tilde n_{c}+\tilde \eta^{c} \tilde \eta_{a} \tilde \eta_{b} \tilde \nabla_{c}(\Omega^{2}\tilde f)\big)\\         
      &=& \big( \tilde \nabla_{a} \tilde n_{b}+ \tilde \eta_{b} \tilde \nabla_{a}(\Omega^{2} \tilde f)+ \tilde \eta_{a}\tilde \nabla_{b}
                             (\Omega^{2} \tilde f)+\tilde \eta^{c} \tilde \eta_{a} \tilde \eta_{b} \tilde \nabla_{c}(\Omega^{2} \tilde f)\big)\\
   &=& \tilde \nabla_{a} \tilde n_{b}+\tilde \eta_{b} \tilde \nabla_{a}( \tilde \eta \cdot \tilde n)
   +\tilde \eta_{a} \tilde \nabla_{b}(\tilde \eta \cdot  \tilde n)+\tilde \eta^{c} \tilde \eta_{a} \tilde \eta_{b} \tilde \nabla_{c}(\tilde \eta \cdot \tilde n).
  \eea
This implies,  
  \bea %
     \tilde \nabla_{a} \tilde \nabla_{b}\Omega & =&  (\tilde \eta \cdot \tilde n) \tilde K_{ab}- \tilde \eta_{b} \tilde \nabla_{a}  
   (\tilde \eta \cdot \tilde n) \label{OmegaEq}
   - \tilde \eta_{a} \tilde \nabla_{b}(\tilde \eta \cdot \tilde n)-\tilde \eta^{c} \tilde \eta_{a}\tilde \eta_{b} \tilde \nabla_{c}(\tilde \eta \cdot \tilde n).
   \eea
   Contracting the indices we get,
 \be \label{boxOmega}
   \tilde  \square \Omega = (\tilde \eta \cdot \tilde n) \tilde K - \tilde \eta^{a}\tilde \nabla_{a} (\tilde \eta \cdot \tilde n).
\ee

Recall that we are working with physical spacetimes satisfying Einstein's equation with positive cosmological constant. For such a spacetime, conformal transformation to unphysical spacetime relate unphysical Ricci scalar and Einstein tensor as follows:
\be
\tilde R = - 2 (d-1) \Omega^{-1} \tilde \square \Omega + d (d-1) \Omega^{-2}(1+\tilde \nabla_{c}\Omega \tilde \nabla^{c}\Omega), \label{R_conf}
\ee
and
\be
\tilde G_{ab}=\Omega^{-1} (d-2) (\tilde g_{ab} \tilde \square  \Omega - \tilde \nabla_{a} \tilde \nabla_{b}\Omega)
    -\frac{\Omega^{-2}}{2} (d-1)(d-2) \tilde g_{ab} \ (1+\tilde \nabla_{c}\Omega \tilde \nabla^{c}\Omega). \label{G_conf}
\ee                                                                                        
Equation \eqref{G_conf} together with
 \eqref{OmegaEq}, \eqref{boxOmega} implies,
\bea
  \label{GeqI}
\tilde h^{c}_{a} \tilde h^{d}_{b} \tilde G_{cd} &=& \Omega^{-1}(d-2)\big\{\tilde h_{ab}(\tilde \eta \cdot \tilde n)\tilde K - \tilde \eta^{c}
\tilde \nabla_{c}(\tilde \eta \cdot \tilde n) \tilde h_{ab} - (\tilde \eta \cdot \tilde n)\tilde K_{ab}\big\} \nn \\ 
 && -\frac{\Omega^{-2}}{2} (d-1)(d-2) \tilde h_{ab}\big(1-(\tilde \eta \cdot \tilde n)^{2}\big).
\eea
From the definition of Einstein tensor we also have,
\bea \label{GeqII}
\tilde G_{cd} \tilde h^{c}_{a} \tilde h^{d}_{b}&=&\tilde R_{ecfd}\tilde h^{c}_{a}\tilde h^{d}_{b} \tilde g^{ef}-\frac{1}{2}\tilde R \tilde h_{ab}\\
&=& \tilde R_{ecfd}\tilde h^{c}_{a} \tilde h^{d}_{b} (-\tilde \eta^{e} \tilde \eta^{f}+\tilde h^{ef})-\frac{1}{2}\tilde R \tilde h_{ab}\\
&=&\tilde R_{ecfd} \tilde h^{c}_{a}\tilde h^{d}_{b} \tilde h^{ef}-(\mathcal{L}_{\tilde \eta}\tilde K_{ab}+\tilde K_{a}^{c}\tilde K_{cb})-\frac{1}{2}\tilde R \tilde h_{ab}\\
&=&\tilde \cR_{ab}-\tilde K_{a}^{c} \tilde K_{cb}+\tilde K \tilde K_{ab}-(\mathcal{L}_{\tilde \eta}\tilde K_{ab}+\tilde K_{a}^{c}\tilde K_{cb}) -\frac{1}{2}\tilde R \tilde h_{ab},
\eea
where in the last steps we have used the Gauss-Codazzi equations. 
Combining \eqref{GeqI} and \eqref{GeqII} and substituting \eqref{R_conf} we get a equation for the unphysical Ricci tensor in terms of unphysical extrinsic curvature,
\bea 
\nn \label{K_New}
\tilde \cR_{ab} &=& 2  \tilde K_{a}^{c} \tilde K_{cb} -\tilde K \tilde K_{ab}+\mathcal{L}_{\tilde \eta}\tilde K_{ab} 
- (d-1)\Omega^{-1}\tilde h_{ab} (\tilde \eta \cdot \tilde n) \tilde K  \\ \nn
& & + \ (d-1)\Omega^{-1} \tilde h_{ab}
\tilde \eta^{c}\tilde \nabla_{c}(\tilde \eta \cdot \tilde n)+ \frac{\Omega^{-2}}{2}d (d-1)\tilde h_{ab}(1-(\tilde \eta \cdot \tilde n)^{2}) \\
& & + \  \Omega^{-1}(d-2)\big\{\tilde h_{ab}(\tilde \eta \cdot \tilde n)\tilde K- \tilde \eta^{c}
\tilde \nabla_{c}(\tilde \eta \cdot \tilde n) \tilde h_{ab} - (\tilde \eta \cdot \tilde n)\tilde K_{ab}\big\} \nn \\
& & 
- \ \frac{\Omega^{-2}}{2} (d-1)(d-2) \tilde h_{ab}\big(1-(\tilde \eta \cdot \tilde n)^{2}\big).
\eea

We would evaluate this equation at the boundary $\Omega=0$. At this point, it is useful to recall definition of function $f$ \eqref{def_f},
\be 
\tilde \eta \cdot \tilde n = \Omega^{2}\tilde f -1, 
\ee
where $\tilde f$ is a smooth function on ${\cal I}^+$, i.e., it admits an expansion in powers of $\O$ as,
\be
 \tilde f=\sum_{n=0}^{\infty}\Omega^{n}\tilde f_{(n)}.
 \ee
 In then follows that, 
 \be
\Omega^{-2}(1-(\tilde \eta \cdot \tilde n)^{2})\big|_{\Omega=0}= 2\tilde f,
 \ee
 and
 \be
   \Omega^{-1}\tilde \eta^{c}\tilde \nabla_{c}(\tilde \eta \cdot \tilde n)\big|_{\Omega=0}=-2\tilde f.
\ee
Using these identities, equation  \eqref{K_New} at the boundary  becomes,
\be
\mathcal{L}_{\tilde \eta } \tilde K_{ab}=\tilde \cR_{ab}-\Omega^{-1}\tilde K \tilde h_{ab}-(d-2)\Omega^{-1}
\tilde K_{ab}-2 \tilde f(d-2)\tilde h_{ab},
\ee
equivalently,
\bea
\Omega^{-1}\tilde K_{ab} &=& \frac{1}{(d-3)} \bigg(\tilde \cR_{ab} - \frac{1}{2(d-2)}\tilde \cR \tilde h_{ab}
\bigg)-\tilde f \tilde h_{ab} \label{rescaled_K}.
\eea
This is the relation used in main text \eqref{rescaled_K_MT}.

\section{Expressing Kerr-de Sitter metric in Fefferman-Graham gauge}

\label{app:examples}

\subsection*{4d}

In this appendix we express Kerr-de Sitter metric in Fefferman-Graham gauge, and explicitly show that it belongs to our phase space. From the asymptotic form of the unphysical metric we read off the electric part of the Weyl tensor via \eqref{FG_4d_MT} and compare it to the expressions given in~\cite{ABKI} by ABK. ABK obtained the electric part of the Weyl tensor by an asymptotic expansion of the Weyl tensor of the unphysical metric. We show that indeed the two expressions are related by expected tensor and conformal transformations. 

The computation below is logically identical to the one presented in the main text for four-dimensional Schwarschild solution, however, detailed expressions are significantly more complicated.

The Kerr-de Sitter metric in Boyer-Lindquist type coordinates take the form~\cite{Carter:1968ks, Akcay:2010vt}\footnote{We use this form of the metric for ease of comparison with the corresponding expressions in~\cite{ABKI}.}
\bea
ds^2 &=& (r^2 + a^2 \cos^2 \theta) \left(\frac{dr^2}{\Delta_r} + \frac{d\theta^2}{1+ \frac{a^2}{l^2} \cos^2 \theta} \right) +  \sin^2 \theta \frac{1+ \frac{a^2}{l^2} \cos^2 \theta}{r^2 + a^2 \cos^2 \theta} \left[ \frac{a dt - (r^2 + a^2) d\phi}{1+\frac{a^2}{l^2}}\right]^2 \nn \\
& & - \frac{\Delta_r}{r^2 + a^2 \cos^2 \theta} \left[\frac{dt - a \sin^2 \theta d \phi}{1+\frac{a^2}{l^2}} \right]^2,
\eea
where
\be
\Delta_r = - \frac{r^4}{l^2} + \left( 1 - \frac{a^2}{l^2}\right) r^2 - 2m r + a^2.
\ee
We express this metric in FG gauge by doing a series of coordinate transformations. To begin with we define a set of coordinate $(\Omega, \chi, \Theta, \bar \phi)$ via the relations,
\bea
t &=& (1+a^2) \log \tan \frac{\chi}{2},\label{t_chi} \\
r &=& \frac{1}{\Omega} \sqrt{1 + a^2 \sin^2 \Theta} \sin \chi \\
 \theta &=& \cos^{-1} \left[ \frac{\cos \Theta}{\sqrt{1 + a^2 \sin^2 \Theta}} \right]  \label{theta_Theta} \\
 \phi &=& \bar \phi + \frac{a}{1+a^2} t, \label{phi_bar_phi} 
\eea
Equation \eqref{theta_Theta} is equivalently written as,
\be
(1 + a^2) \tan^2 \Theta  = \tan^2 \theta.
\ee
At first sight these transformations look complicated, but they are not. The transformations for $\theta$ and $\phi$ are the standard transformations used to manifest asymptotic (anti) de Sitter nature of Kerr-(anti)-de Sitter metrics~\cite{Hawking:1999dp, Papadimitriou:2005ii}. The change from $t$ to $\chi$ is motivated by the Schwarschild example discussed in the main text. In that example it changes the boundary metric from being the  natural metric on $\mathbb{R} \times \mathbb{S}^2$ to the  natural round metric on $\mathbb{S}^3$, upto a conformal factor. The conformal factors are taken into account in the definition of $\Omega$; $\Omega$ is the normalised time coordinate near the boundary with boundary at $\Omega = 0$. 
In these coordinates, the unphysical metric in the requisite form to \emph{zeroth} order
\be
\widetilde{(ds)}^2 = \Omega^2 ds^2 = \tilde g_{\mu \nu} d x^\mu d x^\nu
\ee
with various components
\begin{align}
\tilde g_{\Omega \Omega } &= -1 + \mathcal{O}(\Omega) &
\tilde g_{\Omega \chi } &=  \mathcal{O}(\Omega) \\
\tilde g_{\chi \chi } &= 1 + \mathcal{O}(\Omega) &
\tilde g_{\Omega \Theta } &= \mathcal{O}(\Omega) & \\
\tilde g_{\Theta \Theta } &= \sin^2 \chi + \mathcal{O}(\Omega)& 
\tilde g_{\chi \Theta } &=  \mathcal{O}(\Omega^2) &\\
\tilde g_{\bar \phi \bar \phi } &=  \sin^2 \chi \sin^2 \Theta+ \mathcal{O}(\Omega)&
\tilde g_{\chi \bar \phi } &=  \mathcal{O}(\Omega^3) & \\
\tilde g_{\Theta \bar \phi } &= 0 &
\tilde g_{\Omega \bar \phi } &= 0.
\end{align}
The metric at the boundary is the unit round metric on $\mathbb{S}^3$. To bring this metric in the Fefferman-Graham gauge to the requisite order in $\Omega$,  the following coordinate transformations are required as power series in $\bar \Omega$: 
\bea
\Omega &=& \bar \Omega  + h_1  \bar \Omega^3 + + h_2 \bar \Omega^4 + \ldots  \\
\chi &=&  \bar \chi  + k_1  \bar \Omega^2 +  k_2  \bar \Omega^4 + \ldots  \\
\cos \Theta &=&   \cos \bar \Theta + v_1  \bar \Omega^2 +  v_2  \bar \Omega^4 + \ldots ,
\eea
with functions
\bea
k_1 &=& -\frac{\cot \bar \chi}{2} \\
v_1 &=& \frac{a^2 \sin ^2\bar \Theta \cos \bar \Theta \csc ^2\bar \chi}{2 (1 + a^2 \sin^2 \bar \Theta)}\\
h_1 &=& \frac{1}{4}-  \frac{\csc ^2\bar \chi}{2(1 + a^2 \sin ^2 \bar \Theta)} \\
k_2 &=& -\frac{1}{8} \cot \bar \chi \csc ^2\bar \chi \\
h_2 &=& \frac{1}{3} m \csc ^3\bar \chi \left(1+ a^2 \sin ^2\bar \Theta\right)^{-3/2}
\eea
and
\be
v_2 = -\frac{a^2 \sin ^2\bar \Theta \cos \bar \Theta \csc ^4\bar \chi }{8 \left(1+a^2 \sin ^2\bar \Theta\right)^3}\left(1 -a^2 +5 a^2 \cos
   ^2\bar \Theta - 2 a^4 \sin
   ^4\bar \Theta - \cos 2 \bar \chi \left(1+a^2 \sin ^2\bar \Theta\right)^2\right).
\ee
 We constructed these functions by successively making sure that the metric in the FG gauge at requisite order.  The functions listed above serve the following purposes: 
$k_1$ and $k_2$ ensure that $\tilde g_{\bar\Omega \bar \chi}$ vanishes at orders $\bar \Omega$ and  $\bar \Omega^3$ respectively; $v_1$ and $v_2$  ensure that $\tilde g_{\bar\Omega \bar \Theta}$ vanishes at orders $\bar \Omega$ and  $\bar \Omega^3$ respectively; 
and  finally 
the functions $h_1$ and $h_2$ ensure that $\tilde g_{\bar\Omega\bar\Omega}$ is $-1$ at orders $\bar \Omega^2$ and  $\bar \Omega^3$ respectively.

With these transformations, the Kerr-de Sitter metric is in FG gauge at requisite order,
\bea
\tilde g_{\bar \chi \bar \chi } &=& 1 + \frac{1}{2}\bar \Omega^2 - \frac{2}{3} \bar \Omega^3 \tilde E_{\bar \chi \bar \chi}+ \mathcal{O}(\Omega^4) \\
\tilde g_{\bar \chi \bar \Theta } &=&- \frac{2}{3} \bar \Omega^3 \tilde E_{\bar \chi \bar \Theta } + \mathcal{O}(\Omega^4)\\
\tilde g_{\bar \Theta \bar \Theta } &=& \sin^2 \chi \left( 1+ \frac{1}{2}\bar \Omega^2 \right) - \frac{2}{3} \bar \Omega^3 \tilde E_{\bar \Theta \bar \Theta }+ \mathcal{O}(\Omega^4) \\
\tilde g_{\bar \phi \bar \phi } &= & \sin^2 \chi \sin^2 \bar \Theta \left( 1+ \frac{1}{2}\bar \Omega^2 \right) - \frac{2}{3} \bar \Omega^3 \tilde E_{\bar \phi \bar \phi }+ \mathcal{O}(\Omega^4) 
\eea
together with
\begin{align}
\tilde g_{\bar\Omega \bar  \Omega } &= -1 + \mathcal{O}(\Omega^4) &
\tilde g_{\bar\Omega \bar \chi } &=  \mathcal{O}(\Omega^4)& \\
\tilde g_{\bar\Omega \bar \Theta } &= \mathcal{O}(\Omega^4) &
\tilde g_{\bar \chi \bar \phi } &=  \mathcal{O}(\Omega^4)& \\
\tilde g_{\bar\Omega \bar \phi } &= 0& 
\tilde g_{\bar \Theta \bar \phi } &= 0. & 
\end{align}
The components of the electric part of the Weyl tensor are found to be
\bea
\tilde E_{\bar \chi \bar \chi} &=& - m \ \Sigma^{-5} \ \csc ^3\bar \chi   \
   \left(2- a^2 \sin^2 \bar \Theta \right) \\
\tilde E_{\bar \chi \bar \phi } &=& 3 a m \  \Sigma^{-5}  \ \csc ^2\bar \chi  \ \sin ^2\bar \Theta \  \\
\tilde E_{\bar \Theta \bar \Theta } &=& m \ \Sigma^{-3} \  \csc \bar \chi  \\
\tilde E_{\bar \phi \bar \phi }&=&m  \ \Sigma^{-5} \ \sin ^2\bar \Theta \  \csc \bar \chi  \ \left(1 -2 a^2 \sin^2 \bar \Theta \right),
\eea
where
\be
\Sigma =  \sqrt{1+ a^2 \sin^2\bar \Theta}.
\ee

We can now compare these expressions with those in reference~\cite{ABKI}.
There the boundary metric is taken to 
\be
ds^2_{\mathcal{I}^+} = \frac{dt^2}{(1+a^2)^2} - \frac{2 a \sin^2 \theta d t d\phi }{\left(1+a^2\right)^2} + \frac{ d \theta^2}{{1+a^2\cos^2 \theta}} + \frac{\sin^2 \theta d\phi^2}{1+a^2}, \label{metric_boundary}
\ee
and the electric part of the Weyl tensor is written to be
\begin{align}
\tilde E_{tt}^\rom{ABK} &= - \frac{2m}{(1+a^2)^2},&\\
\tilde E_{t\phi}^\rom{ABK} &= \frac{2 a m}{(1+a^2)^2} \sin^2 \theta,& \\
\tilde E_{\theta \theta}^\rom{ABK} &=  \frac{m}{(1+a^2 \cos^2 \theta)},&\\
\tilde E_{\phi\phi}^\rom{ABK} &= \frac{m}{(1+a^2)^2} \sin^2 \theta \left(1-a ^2 \left(\frac{1}{2} - \frac{3}{2} \cos 2 \theta \right)\right).&
\end{align}
Under the diffeomorphism (cf. equations \eqref{t_chi}, \eqref{phi_bar_phi}, \eqref{theta_Theta}): 
\begin{align}
t &= (1+a^2) \log \tan \frac{\bar \chi}{2}, &
\cos \theta &= \frac{\cos \bar \Theta}{\Sigma} &
 \phi &= \bar \phi + \frac{a}{1+a^2} t, &
\end{align}
the boundary metric \eqref{metric_boundary} becomes conformal to the round metric on $\mathbb{S}^3$ \be
ds^2_{\mathbb{S}^3} = \bar \omega^2 ds^2_{\mathcal{I}^+}   
\ee
with the conformal factor $\bar \omega  = \sin \bar \chi \ \Sigma.$
Under these transformations, one can indeed verify with a direct calculation that 
$
\tilde E_{ab} = (\bar \omega)^{-1}\ \tilde E_{ab}^\rom{ABK}.
$
Conserved charges for Kerr-de Sitter metric are discussed by ABK.

\subsection*{5d}

We now express the five-dimensional Myers-Perry-de Sitter metric in Fefferman-Graham gauge. This computation is similar to the one presented above for four-dimensional Kerr-de Sitter metric, if somewhat more involved. We explicitly verify that the five-dimensional Myers-Perry-de Sitter metric belongs to our phase space. From the asymptotic form of the metric we read off the electric part of the Weyl tensor via \eqref{FG_4d_MT}. 

To set the conventions, let us recall that Einstein's equations with positive cosmological constant $\Lambda$ are
$
R_{ab} - \frac{1}{2} g_{ab} R + \Lambda g_{ab} = 0,
$ with de Sitter length $\ell$ defined as
$
\Lambda = \frac{1}{2\ell^2} (d-1)(d-2).
$
These equations simplify to
\be
R_{ab} = \frac{(d-1)}{\ell^2} g_{ab}. \label{einstein_eqs}
\ee
In odd $d=2n+1$ dimensions Kerr-de Sitter metrics 
 satisfying Einstein equation \eqref{einstein_eqs}
take the following form~\cite{Gibbons:2004uw} in Boyer-Lindquist coordinates (with $\mu_i$s as direction cosines and $\lambda = \ell^{-2}$):
\bea
ds^2 &=& - W \, (1 -\lambda r^2)\, 
dt^2 + \fft{U \, dr^2}{V-2m} +
\fft{2m}{U}\Big(dt - \sum_{i=1}^n \fft{a_i\, \mu_i^2\, d\varphi_i}{
1 + \lambda\, a_i^2}\Big)^2 \nn\\
&& 
+ \sum_{i=1}^n \fft{r^2 + a_i^2}{1 + \lambda\, a_i^2}
\, [d\mu_i^2 + \mu_i^2\, (d\varphi_i -\lambda\, a_i\, dt)^2] \nn\\
&&+
\fft{\lambda}{W\, (1-\lambda r^2)}
\Big( \sum_{i=1}^n \fft{(r^2 + a_i^2)\mu_i\, d\mu_i}{
1 + \lambda\, a_i^2}\Big)^2 \,,\label{BL}
\eea
where
\begin{align}
W &\equiv \sum_{i=1}^n \fft{\mu_i^2}{1+\lambda\, a_i^2}\, &
F& \equiv \fft{r^2}{1-\lambda\, r^2}\, \, 
  \sum_{i=1}^n \fft{\mu_i^2}{r^2+a_i^2},& \\
U &\equiv \sum_{i=1}^n \fft{\mu_i^2}{r^2 + a_i^2}\, 
\prod_{j=1}^n (r^2 + a_j^2), &
V &\equiv \fft1{r^2}\, (1-\lambda\, r^2)\, \prod_{i=1}^n (r^2 + a_i^2)
= \fft{U}{F}\,,
\end{align}
Since we are interested in five-dimensions, we set $n=2$. In addition we use the notation
\begin{align}
\mu_1 &= \sin \theta, & \mu_2 &= \cos \theta, &
a_1 &= a, &  a_2 &= b, & 
\varphi_1 &= \phi, & \varphi_2 &= \psi. &
\end{align}
The metric  in $(t, r, \theta, \phi, \psi)$ coordinates simplifies to the following more standard form
\bea\label{metric_MP}
ds^2 & = & -\frac{V-2m}{U}\left(dt-\frac{a\sin^2\theta}{\Xi_a}d\phi
-\frac{b\cos^2\theta}{\Xi_b}d\psi\right)^2+
\frac{\Delta_\th \sin^2\th}{U}\left(a dt-\frac{r^2+a^2}{\Xi_a}d\phi\right)^2\nn\\
&&+\frac{\Delta_\th \cos^2\th}{U}\left(bdt-\frac{r^2+b^2}{\Xi_b}d\psi\right)^2+
\frac{U}{V-2m}dr^2+\frac{U}{\Delta_\theta}d\th^2\nn\\
&&+\frac{(1-\lambda r^2)}{r^2U}\left(a b dt-\frac{b(r^2+a^2)\sin^2\theta}{\Xi_a}
d\phi-\frac{a(r^2+b^2)\cos^2\theta}{\Xi_b}d\psi\right)^2,
\eea
where
\begin{align}
U &= r^2+a^2\cos^2\theta+b^2\sin^2\theta,& 
V  &= \frac{1}{r^2}(1- \lambda r^2)(r^2+a^2)(r^2+b^2),&\\
\Delta_\theta &= W \ \Xi_a \  \Xi_b = 1+ \lambda a^2 \cos^2\th + \lambda b^2 \sin^2\th,&
\Xi_a& =1 + \lambda a^2,\qquad \Xi_b=1+ \lambda b^2.&
\end{align}

From now on we set de Sitter length to unity $\ell = 1.$ To begin with we introduce $\Omega = 1/r$ and choose the conformal factor $\Omega^2$ to obtain the unphysical metric,
\be
\widetilde{(ds)}^2 = \Omega^2 ds^2 = \tilde g_{\mu \nu} d x^\mu d x^\nu.
\ee  
To leading order in $\Omega$, $\tilde g_{\Omega \Omega} = 1$ and the intrinsic four-dimensional metric at the  boundary ${\mathcal{I}^{+}}$  is:
\bea
ds^2_{\mathcal{I}^{+}}= dt^2 + \frac{d\theta^2}{\Delta_\theta} + \frac{\sin^2 \theta}{\Xi_a}d\phi(d\phi - 2 a dt)+ \frac{\cos^2 \theta}{\Xi_b}d\psi(d\psi - 2 b dt). \label{boundary_metric_5d}
\eea
This metric is conformal to the round metric on $\mathbb{S}^4$, which can be made explicit as follows. Define new coordinates $(\chi,  \Theta, \bar \phi, \bar \psi)$ via~\cite{Hawking:1999dp, Papadimitriou:2005ii}:
\bea
 \phi &=& \bar \phi + a t, \label{transformations} \\
 \psi &=&\bar \psi +  b t, \\
  \theta &=& \cos^{-1}\left[\frac{\sqrt{1+b^2} \cos  \Theta}{\sqrt{1 + a^2 \sin^2  \Theta+b^2  \cos^2  \Theta}}\right],\\
 t &=& \log \tan \frac{\chi}{2}, 
\eea
the boundary metric becomes
\be
ds^2_{\mathbb{S}^{4}} = \omega^2 ds^2_{\mathcal{I}^{+}}, \label{conformal}
\ee
where
\be
\omega = \sin \chi \sqrt{1 + a^2 \sin^2  \Theta+b^2  \cos^2  \Theta},
\ee
and 
\be
ds^2_{\mathbb{S}^{4}} = d\chi^2 + \sin^2 \chi (d\Theta^2 + \sin^2 \Theta \ d\bar \phi^2 + \cos^2 \Theta  \ d\bar \psi^2).
\ee
The electric part of the Weyl tensor,
\be
(\tilde E_{ac})_0 = \frac{1}{2} \lim_{\Omega \to 0} \Omega^{-2} (\tilde C_{abcd} n^b n^d),
\ee 
for the unphysical metric in $(t, \Omega, \theta, \phi, \psi)$ coordinates can be easily computed with the help of Mathematica. We call this tensor $\tilde E^\rom{ABK}_{ab}$ as it is the five-dimensional  analog of the ABK Kerr-de Sitter expressions:
\begin{align}
\tilde E^\rom{ABK}_{tt} &=  - 3m ,&
\tilde E^\rom{ABK}_{\theta \theta}& = \frac{m}{1 + a^2 \cos^2 \theta +b^2 \sin^2 \theta},&\\
\tilde E^\rom{ABK}_{\phi \phi}&= \frac{m \sin^2\theta}{(1+ a^2)^2}( 1 - a^2 + 2 a^2 \cos2  \theta),&
\tilde E^\rom{ABK}_{\psi \psi}&=\frac{m \cos^2\theta}{(1+ b^2)^2}( 1 - b^2 -2 b^2 \cos2  \theta),&\\
\tilde E^\rom{ABK}_{t\phi}&=\frac{3  ma \sin ^2\theta }{1+a^2},&
\tilde E^\rom{ABK}_{t\psi}&=\frac{3 mb \cos ^2\theta }{1+b^2}.&
\end{align}

Now we do a series of coordinate transformations to go from $(\Omega, \chi, \Theta, \bar \phi, \bar \psi)$ to  $(\bar \Omega, \bar \chi, \bar \Theta, \bar \phi, \bar \psi)$ such that the unphysical Myers-Perry-de Sitter metric is in the Fefferman-Graham gauge to requisite order. To begin with we have
\begin{align}
\tilde g_{\Omega \Omega } &= -1 + \mathcal{O}(\Omega) &
\tilde g_{\Omega \chi } &=  \mathcal{O}(\Omega) & 
\tilde g_{\Omega \Theta } &= \mathcal{O}(\Omega) & \\
\tilde g_{\Omega \bar \phi } &= 0 & 
\tilde g_{\Omega \bar \psi } &= 0 &
\tilde g_{\chi \chi } &= 1 + \mathcal{O}(\Omega^2) & \\
\tilde g_{\chi \Theta } &=  \mathcal{O}(\Omega^2) &
\tilde g_{\chi \bar \phi } &=  \mathcal{O}(\Omega^4) & 
\tilde g_{\chi \bar \psi } &=  \mathcal{O}(\Omega^4) & \\
\tilde g_{\Theta \Theta } &= \sin^2 \chi + \mathcal{O}(\Omega^2) &
\tilde g_{\Theta \bar \phi } &= 0 & 
\tilde g_{\Theta \bar \psi } &= 0 & \\
\tilde g_{\bar \phi \bar \phi } &=  \sin^2 \chi \sin^2 \Theta+ \mathcal{O}(\Omega^2)&
\tilde g_{\bar \phi \bar \psi } &=  \mathcal{O}(\Omega^4)&
\tilde g_{\bar \psi \bar \psi } &=  \cos^2 \chi \sin^2 \Theta+ \mathcal{O}(\Omega^2).&
\end{align}

The metric at the boundary is now round metric on $\mathbb{S}^4$. We change
\bea
\Omega &=& \bar \Omega  + h_1  \bar \Omega^3 + + h_2 \bar \Omega^5 + \ldots  \\
\chi &=&  \bar \chi  + k_1  \bar \Omega^2 +  k_2  \bar \Omega^4 + \ldots  \\
\cos \Theta &=&   \cos \bar \Theta + v_1  \bar \Omega^2 +  v_2  \bar \Omega^4 + \ldots ,
\eea
with appropriate functions successively constructed to  ensure that the metric in the FG gauge at requisite order.  These functions serve the following purposes: 
$k_1$ and $k_2$ ensure that $\tilde g_{\bar\Omega \bar \chi}$ vanishes at orders $\bar \Omega$ and  $\bar \Omega^3$ respectively; $v_1$ and $v_2$  ensure that $\tilde g_{\bar\Omega \bar \theta}$ vanishes at orders $\bar \Omega^2$ and  $\bar \Omega^4$ respectively; 
and  finally 
the functions $h_1$ and $h_2$ ensure that $\tilde g_{\bar\Omega\bar\Omega}$ is $-1$ at orders $\bar \Omega^2$ and  $\bar \Omega^4$ respectively. Expressions for these functions are omitted, as some of them are exceedingly long, and not illuminating.

With these transformations, the five-dimensional Myers-Perry-de Sitter metric is in FG gauge at requisite order, i.e.,
\begin{align}
\tilde g_{\bar\Omega \bar  \Omega } &= -1 + \mathcal{O}(\Omega^5), &
\tilde g_{\bar\Omega \bar \chi } &=  \mathcal{O}(\Omega^5),& \\
\tilde g_{\bar\Omega \bar \Theta } &= \mathcal{O}(\Omega^5), & 
\tilde g_{\bar\Omega \bar \phi } &= \mathcal{O}(\Omega^5), & \\
\tilde g_{\bar\Omega \bar \psi } &= \mathcal{O}(\Omega^5), & 
\tilde g_{\bar\chi \bar \Theta } &= \mathcal{O}(\Omega^6), & \\
\tilde g_{\bar \Theta \bar \phi } &= 0,  & 
\tilde g_{\bar \Theta \bar \phi } &= 0.  &
\end{align}
together with the remaining components of the form
\be
\tilde g_{ab} = - \nabla_a \Omega \nabla_b \Omega + \left( 1+ \frac{1}{2} \Omega^2 + \frac{1}{16}\Omega^4 \right)  (\tilde h_{ab})_0  - \frac{1}{2} \Omega^4 (\tilde E_{ab})_0 + \mathcal{O}(\Omega^{5}),
\ee
with the electric part of Weyl tensor $(\tilde E_{ab})_0$
\bea
\tilde E_{\bar \chi \bar \chi} &=& m \ \csc^4\bar \chi \ \Sigma^{-3} (\Sigma -4),\\
\tilde E_{\bar \chi \bar \phi} &=&  4 m a \ \sin^2\bar \Theta \ \csc^3\bar \chi \ \Sigma^{-3}, \\
\tilde E_{\bar \chi \bar \psi} &=& 4 m b \ \cos^2\bar \Theta \ \csc^3\bar \chi \ \Sigma^{-3},\\
\tilde E_{\bar \Theta \bar \Theta} &=&  m   \csc^2\bar \chi \ \Sigma^{-2}, \\
\tilde E_{\bar \phi \bar \phi} &=& m   \csc^2\bar \chi \ \sin^2 \bar \Theta \Sigma^{-3}  \ (\Sigma - 4 a^2 \sin^2 \bar \Theta),\\
\tilde E_{\bar \psi \bar \psi} &=& m   \csc^2\bar \chi \ \cos^2 \bar \Theta \Sigma^{-3}  \ (\Sigma - 4 b^2 \cos^2 \bar \Theta), \\
\tilde E_{\bar \phi \bar \psi} &=& -4 m a b  \csc^2\bar \chi \ \sin^2 \bar \Theta  \cos^2 \bar \Theta \ \Sigma^{-3} ,
\eea
where
\be
\Sigma =  (1 + a^2 \sin^2 \bar \Theta + b^2 \cos^2 \bar \Theta).
\ee

Under the change of coordinates and the conformal frame that takes boundary metric from \eqref{boundary_metric_5d} to unit round metric on the four-sphere, cf.~\eqref{transformations}--\eqref{conformal}, 
one can indeed verify with a direct calculation that,
$
\tilde E_{ab} = \omega^{-2}\ \tilde E_{ab}^\rom{ABK}.
$

\end{document}